\documentclass[journal]{IEEEtran}

\usepackage{balance}
\usepackage[ruled]{algorithm2e}
\usepackage{amssymb}
\usepackage{mathbbol}
\usepackage{stmaryrd}
\usepackage{cite}
\usepackage{color}
\usepackage{multirow}
\usepackage{subfigure}
\usepackage{graphicx,times,amsmath}
\usepackage{epstopdf}
\usepackage{pifont}

\ifCLASSINFOpdf

\else

\fi

\begin{document}
\title{Cross-Technology Communications for Heterogeneous IoT Devices Through Artificial Doppler Shifts}

\author{Wei~Wang,
	Shiyue~He,
	Liang~Sun,
	Tao~Jiang,~\IEEEmembership{Fellow,~IEEE},
	Qian~Zhang,~\IEEEmembership{Fellow,~IEEE}
	\thanks{The work was supported in part by the National Science Foundation of China with Grant 61871441, 91738202, 61729101, 61671003, Major Program of National Natural Science Foundation of Hubei in China with Grant 2016CFA009, the RGC under Contract CERG 16203215, Contract ITS/143/16FP-A, the Guangdong Natural Science Foundation No. 2017A030312008 and the Key Laboratory of Dynamic Cognitive System of Electromagnetic Spectrum Space (Nanjing Univ. Aeronaut. Astronaut.), Ministry of Industry and Information Technology, Nanjing, 211106, China with KF20181911, the Fundamental Research Funds for the Central Universities under Grant YWF-18-BJ-Y-176.}

	\thanks{W. Wang, and T. Jiang are with the School of Electronic Information and Communications and the Wuhan National Laboratory for Optielectronics, Huazhong University of Science and Technology. E-mail: \{weiwangw,taojiang\}@hust.edu.cn.}
	\thanks{S. He is with the School of Electronic Information and Communications, Huazhong University of Science and Technology and the Key Laboratory of Dynamic Cognitive System of Electromagnetic Spectrum Space (Nanjing Univ. Aeronaut. Astronaut.), Manistry of Industry and Information Technology. Email: \{shiue\_he\}@hust.edu.cn.}
	\thanks{L. Sun is with Beihang University, China. E-mail: {eelsun@buaa.edu.cn.}, and is the corresponding author.}
	\thanks{Q. Zhang is with the Department of Computer Science and Engineering, Hong Kong University of Science and Technology.  E-mail: qianzh@cse.ust.hk.}}

\maketitle
\begin{abstract}
Recent years have seen major innovations in developing energy-efficient wireless technologies such as Bluetooth Low Energy (BLE) for Internet of Things (IoT). Despite demonstrating significant benefits in providing low power transmission and massive connectivity, hardly any of these technologies have made it to directly connect to the Internet. Recent advances demonstrate the viability of direct communication among heterogeneous IoT devices with incompatible physical (PHY) layers. These techniques, however, require modifications in transmission power or time, which may affect the media access control (MAC) layer behaviors in legacy networks. In this paper, we argue that the frequency domain can serve as a free side channel with minimal interruptions to legacy networks. To this end, we propose DopplerFi, a communication framework that enables a two-way communication channel between BLE and Wi-Fi by injecting artificial Doppler shifts, which can be decoded by sensing the patterns in the Gaussian frequency shift keying (GFSK) demodulator and Channel State Information (CSI). The artificial Doppler shifts can be compensated by the inherent frequency synchronization module and thus have a negligible impact on legacy communications. Our evaluation using commercial off-the-shelf (COTS) BLE chips and 802.11-compliant testbeds have demonstrated that DopplerFi can achieve throughput up to 6.5~Kbps at the cost of merely less than 0.8\% throughput loss.
\end{abstract}

\maketitle

\section{Introduction}

The wide deployments of Wi-Fi and cellular network infrastructures have provided ubiquitous and transparent access to the Internet for today's laptops, tablets, and smartphones. The coming wave of Internet of Things (IoT), however, does not enjoy the same level of ubiquity in Internet access. This is because these tiny, low-end IoT devices tend to adopt low-power wireless technologies instead of power-intensive Wi-Fi or cellular technologies~\cite{zachariah2015internet}. As a mature technology, Bluetooth Low Energy (BLE) is widely adopted in today's wearables and smart objects, and is envisioned to continue its dominance in the market in near future~\cite{ABI}. These BLE-based IoT devices connect to the Internet through a gateway, which can be a smartphone or an access point equipped with both BLE and Wi-Fi/Ethernet interfaces. Although a gateway can bring Internet connectivity to IoT devices using protocols that are incompatible to Wi-Fi, it induces a large amount of traffic overhead and requires additional gateway deployments, which are not scalable to massive IoT connectivity~\cite{b2w2}.

Recent research efforts have been devoted to enable direct communications between heterogeneous IoT protocols and Wi-Fi. These pioneering designs enable cross-technology communications (CTC) by embedding bits into transmission power levels~\cite{b2w2,wizig} or packet transmission time shifts/patterns~\cite{freebee,cmorse,DCTC,EMF}. Despite that these innovations are transparent to legacy links in that they do not modify frame format or introduce extra packets, they still affect the media access control (MAC) layer behaviors in existing networks. Transmission power modifications induce changes in interference and communication ranges. Perturbing packet transmission time, though merely inducing negligible latency at application layer, may disturb contention behaviors.

Instead of relying on the amplitude and time dimensions that easily affect MAC behaviors, we argue that a more controllable and non-intrusive way is to exploit the frequency domain. Our observation is that today's communication technologies are robust to carrier frequency perturbations. Such a capability lies in the need to combat against carrier frequency offset (CFO) caused by the Doppler effect and hardware impairments. The current designs of BLE and Wi-Fi can tolerate and compensate up to 150~KHz CFO, which is much higher than the inherent CFO. It leaves enough redundancy to encode bits in the carrier frequency. In addition, such an amount of CFO has little impact on adjacent channels due to the guard band protection.

In this paper, we propose \texttt{DopplerFi}, which aims to enable two-way cross-technology communications between Wi-Fi and BLE by subtly shifting the sender's carrier frequency. DopplerFi is inspired by the Doppler effects caused by movements in legacy networks. It intentionally injects artificial Doppler shifts in legacy packets. These artificial Doppler shifts hide themselves in inherent carrier frequency perturbations and are transparent to MAC and upper layers. Artificial Doppler shifts are injected into the transmitted packets, which can be easily realized by the carrier frequency calibration capability of BLE and Wi-Fi radios. 
Injecting a controllable amount of frequency shift induces minimal impact on existing transmissions as legacy physical layers (PHY) have already been designed to eliminate the impact of CFO.

Translating the above idea into a practical system, however, entails a variety of challenges. Although BLE and Wi-Fi receivers are able to detect CFO from PHY-compatible packets, there is no PHY module that can directly identify frequency shifts from incompatible packets. The first hurdle comes from the fact that BLE adopts much narrower channel compared to today's Wi-Fi. 
Wi-Fi with orthogonal frequency division multiplexing (OFDM) modulation (e.g., IEEE 802.11ac/n/g/a) uses 20 MHz channels while today's BLE channels are 1~MHz or 2~MHz wide. Even if Wi-Fi packets are shifted by different amount of frequencies (in an order of KHz), one overlapped BLE channel would still be overwhelmingly filled by Wi-Fi signals in the frequency domain. Thus, BLE cannot detect the frequency shifts in Wi-Fi packets by simply looking at the frequency domain. To overcome this hurdle, our fundamental insight is that all Wi-Fi packets prepend the same preamble whose frequency patterns can be recognized at a granularity of one subcarrier. The bandwidth of one subcarrier in Wi-Fi is 312.5~KHz, and thus the shifts in such a bandwidth can be captured by BLE radios. To extract the preamble frequency patterns using standard BLE chips, we cannot directly analyze the frequency domain as there is no module in BLE such as fast Fourier transform (FFT) that can obtain the frequency domain signals. Our observation is that the preamble frequency patterns can be reflected in the output of the standard Gaussian frequency shift keying (GFSK) demodulator in BLE. As such, we extract the output bits of GFSK demodulator from BLE's PHY, and then decode the frequency shifts injected into the Wi-Fi packets.

Another challenge stems from decoding BLE frequency shifts using the standard PHY of Wi-Fi receivers. Although there is a module in Wi-Fi reception pipeline to estimate and compensate CFO, it requires a Wi-Fi preamble which is not possessed by a standard BLE packet. Instead, we extract frequency shifts in BLE packets by analyzing the channel state information (CSI), which can be extracted from commercial off-the-shelf (COTS) Wi-Fi cards. Since one BLE channel overlaps with multiple adjacent subcarriers in Wi-Fi, different amounts of frequency shifts in BLE packets can be differentiated by analyzing the frequency correlations among adjacent CSI values.

We implement DopplerFi on TI CC2400 BLE devices~\cite{cc2400}, and WARP~\cite{warpProject}. Evaluation results of validated DopplerFi in creating a reliable two-way free side channel between BLE and Wi-Fi under a wide range of scenarios. DopplerFi achieves throughput up to 6.5~Kbps in an interference-free environment and 1.59~Kbps in a crowded environment with 20+ Wi-Fi access points (APs). On the other hand, DopplerFi induces merely 0.8\% and 0.3\% throughput loss on legacy Wi-Fi and BLE links, respectively.

The contributions of this paper are summarized as follows.
\begin{itemize}
	\item We provide a comprehensive study toward creating a free side channel between heterogeneous IoT devices and Wi-Fi devices in the frequency. The frequency domain side channel imposes minimal impact on MAC behaviors in legacy networks.
	\item Our solution requires no hardware or PHY changes and decodes cross-technology bits by extracting patterns from the inherent GFSK and CSI readings that are readily available in the standard PHYs. 
	\item We present BLE and Wi-Fi prototype implementations on TI CC2400 BLE devices and WARP, and validate the performance under various environments.
\end{itemize}

The remainder of this paper is structured as follows. We begin in Section~\ref{sec:cfo} with an introduction of carrier frequency shifts and our motivation. We elaborate the detailed system design in Section~\ref{sec:design}, followed by our system implementation and evaluation in Section~\ref{sec:evaluate}. Related work is reviewed in Section~\ref{sec:literature}, followed by the discussion in Section~\ref{sec:discuss}. Section~\ref{sec:conclude} concludes this work.

\section{Exploiting Carrier Frequency Shifts}
\label{sec:cfo}

In a typical wireless communication system, the CFO is normally within a limited range of 500~Hz~\cite{fica}. A standard OFDM-based Wi-Fi receiver estimates and compensates CFO using both short and long training symbols in the preamble. Conventional OFDM receivers use both for coarse and fine frequency synchronization, and then compensate the offset to eliminate its impact on the following data symbols. Theoretically, the maximum recoverable CFO is 625~KHz when the two-stage CFO compensation algorithm is employed at the receiver~\cite{heiskala2001ofdm} for a 20~MHz Wi-Fi link. According to 802.11 standards~\cite{ieee2010ieee}, the transmitted center frequency tolerance shall be $\pm$~20~ppm maximum. The maximum tolerable CFO regulated by the 802.11 standard is 232~KHz while the CFO caused by the oscillator should not exceed 96~KHz. Hence, it still complies with the regulation of 802.11 and does not affect the legacy transmission by injecting frequency shifts of less than 136~KHz.

Departure from Wi-Fi, BLE employs frequency hopping techniques with the carrier modulation using GFSK. BLE is much more robust to CFO errors. The maximum tolerable frequency offset error is 150~KHz for a 2~MHz BLE link~\cite{bluetooth2010bluetooth}. Some BLE chips incorporate the feature of crystal drift compensation to match the central frequency of the transmitted signal. With such a feature, the tolerable CFO can be related to around 250~KHz~\cite{cc2400}.

\begin{figure}[t]
	\center
	\includegraphics[width=0.35\textwidth]{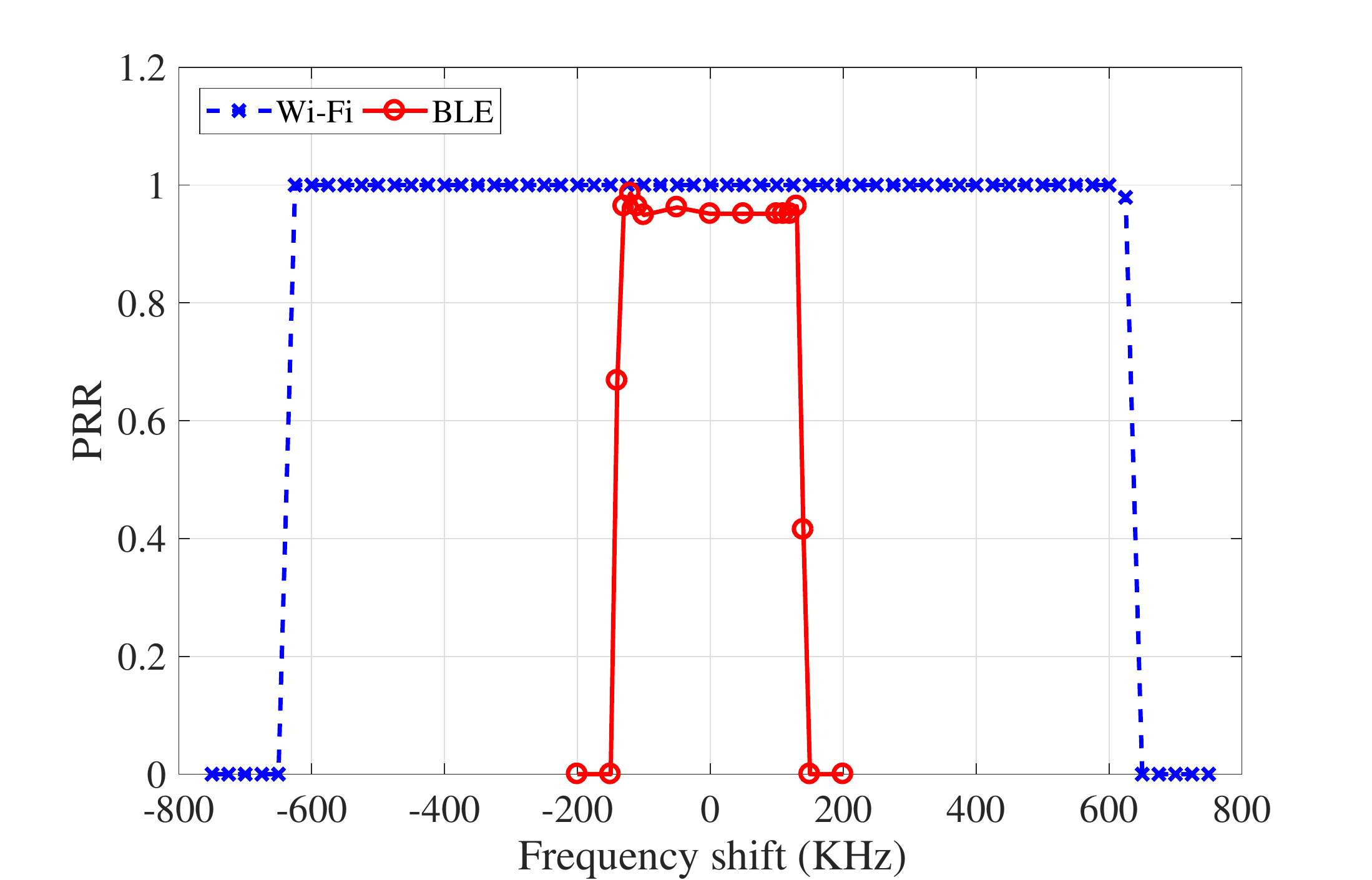}\vspace{0.2cm}
	\caption{Impact of frequency shift on legacy transmissions.}
	\label{fig:feasibility}\vspace{0.3cm}
\end{figure}

Therefore, we observe that there exists a sufficient amount of redundancy that allows legacy radios to freely modulate their carrier frequencies. Fig.~\ref{fig:feasibility} sketches the simulation and experiment results, which clearly show that tuning the carrier frequency within a proper range ((-625~KHz, 625~KHz) for Wi-Fi and (-130~KHz, 130~KHz) for BLE) has a negligibly small impact on legacy transmissions. Additionally, we also test the interference to adjacent channels and observe that the frequency shift within 200~KHz does not affect the performance of other links due to the 2~MHz guard band or channel interval in both Wi-Fi and BLE. These observations imply that with careful design, we can perturb the carrier frequency to create a free channel between Wi-Fi and BLE.

\section{DopplerFi Design}
\label{sec:design}

This section describes an overview of DopplerFi, followed by design specifics in each component. Since Bluetooth will continue its dominance in the smart IoT market~\cite{ABI}, we use the detailed design of cross-technology communications between BLE and Wi-Fi to demonstrate the idea of DopplerFi.

\subsection{Design Overview} 

DopplerFi facilitates BLE-enabled IoT devices to directly communicate with Wi-Fi devices without strings. Simply put, DopplerFi runs a lightweight signal processing block sitting between PHY and MAC. DopplerFi allows direct communication between heterogeneous radio simultaneously with legacy BLE and Wi-Fi transmissions. It injects CFO without interrupting BLE and Wi-Fi links, thus has minimal impact on their network protocols and behaviors. It can be supported by existing Wi-Fi and BLE radios and protocols. DopplerFi does not modify the MAC-related configurations such as transmission power or time, and thus is transparent to upper layers.

Architecturally, DopplerFi is similar to a single frame-level control function such as rate adaptation or power control, in that it merely tunes one parameter, i.e., carrier frequency, in PHY configurations. Departure from rate adaptation and power control functions, DopplerFi works in a cross-layer fashion to pass messages between PHY and upper layers. In the sender mode, DopplerFi tunes the carrier frequency by CFO calibration in PHY. In the receiver mode, DopplerFi extracts CSI values or GFSK outputs from PHY interfaces. 

\begin{figure}[t]
	\center
	\includegraphics[width=0.4\textwidth]{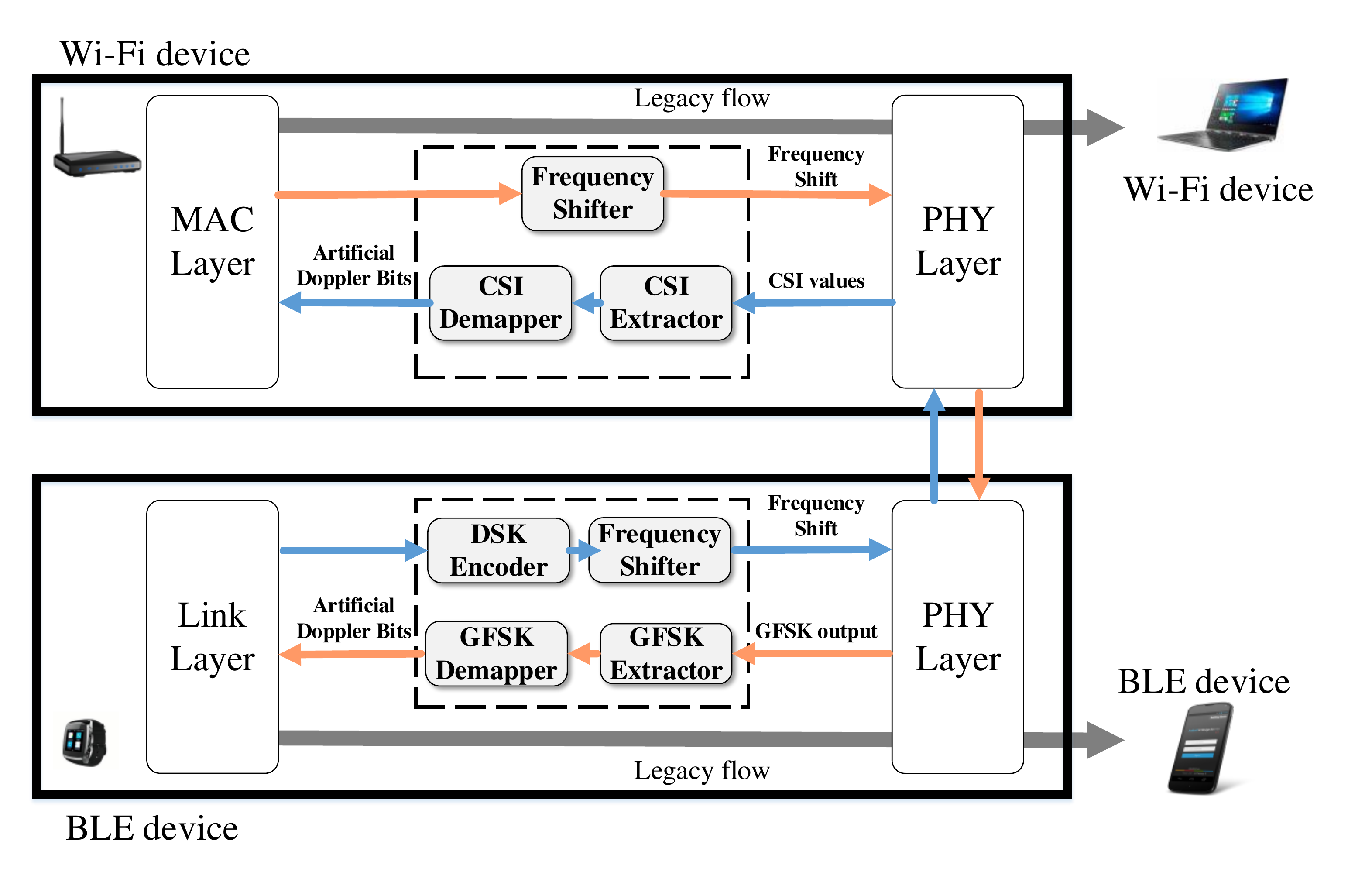}\vspace{0.2cm}
	\caption{Overview of DopplerFi.}
	\label{fig:architecture}\vspace{0.3cm}
\end{figure}

Fig.~\ref{fig:architecture} gives an overview of the DopplerFi architecture in Wi-Fi and BLE transceivers. DopplerFi extends legacy Wi-Fi and BLE by adding the following components.

\begin{itemize}
	\item In the BLE transceiver, DopplerFi contains four modules: the \textit{DSK (Doppler Shift Keying) encoder} and the \textit{frequency shifter} that work in the sender mode, and the \textit{GFSK extractor} and the \textit{GFSK demapper} that work in the receiver mode. The DSK encoder reads the frequency hopping pseudo random sequence in BLE and determines when to embed bits into carrier frequency. The frequency shifter converts bits into carrier frequency shifts. As such, DopplerFi modulates side-channel bits into the carrier frequency of transmitted BLE packets. The GFSK extractor extracts output bits from the inherent GFSK demodulator in the BLE PHY. The GFSK demapper demodulates frequency shifts based on the patterns in the GFSK output bit sequence.

	\item In the Wi-Fi transceiver, DopplerFi contains three modules: the \textit{frequency shifter} that work in the sender mode, and the \textit{CSI extractor} and the \textit{CSI demapper} that work in the receiver mode. Analogous to the BLE transceiver, the Wi-Fi transceiver subtly varies its carrier frequency according to the frequency shifter. In contrast to GFSK-based demodulation in BLE, the Wi-Fi transceiver extracts embedded side-channel bits by extracting CSI using the CSI extractor. The CSI values are fed to the CSI demapper to demodulate the embedded bits by analyzing the variance in the CSI readings caused by BLE frequency shifts.
\end{itemize}

The components in DopplerFi enable bidirectional communications between Wi-Fi and BLE while retaining ongoing legacy BLE and Wi-Fi transmissions.

\subsection{Embedding Artificial Doppler Shifts}

Wi-Fi sends packets in fixed channels with pre-defined carrier frequencies. Although BLE utilizes frequency-hopping spread spectrum (FHSS) technology to change the transmission channel between adjacent packets, the carrier frequency of each channel is fixed. DopplerFi establishes a cross-technology channel by embedding symbols within these carrier frequencies. To shift carrier frequency without affecting Wi-Fi and BLE legacy transmissions, the following requirements should be satisfied. First, the amount of frequency shift should not exceed the limit of recoverable CFO. Additionally, the injected frequency shifts should be incorporated with the frequency hopping in BLE without any modification to the FHSS mechanism. We refer the frequency shift satisfying the above requirements as \textit{artificial Doppler shift}, implying that it has minimal impact on existing systems just like Doppler shifts caused by normal movements.

To meet the above requirements, DopplerFi carefully selects different amounts of frequency shifts for BLE and Wi-Fi respectively, and employs a special companion referred to as the DSK encoder to modulate BLE carrier frequencies in conjunction with FHSS.

\begin{figure}[t]
	\center
	\includegraphics[width=0.5\textwidth]{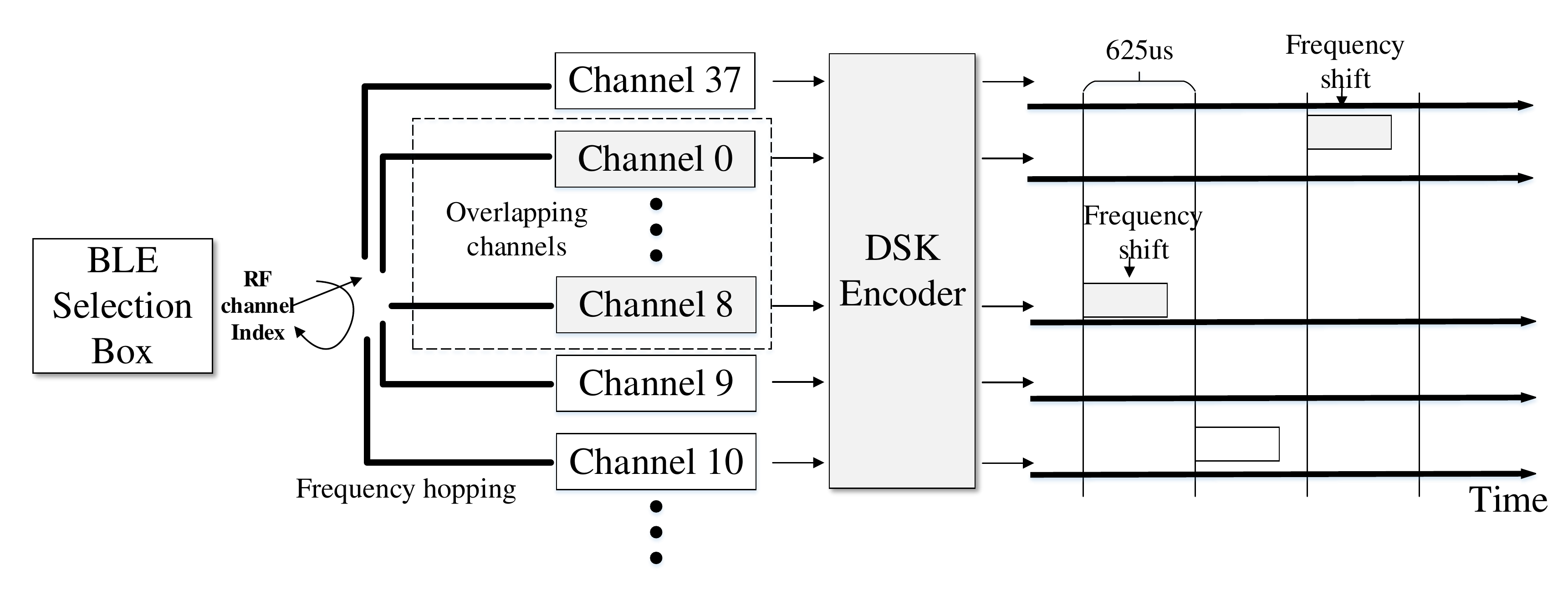}\vspace{0.2cm}
	\caption{DSK encoder.}
	\label{fig:dsk}\vspace{0.1cm}
\end{figure}{}

In standard Wi-Fi PHY, the build-in CFO compensation module can correct frequency offset up to 625~KHz, when the receiver employs two-stage CFO compensation.
To deliver as many side channel bits as possible in each packet, one might use multiple levels of frequency shifts for modulation. Unfortunately, since BLE adopts 1~MHz or 2~MHz bandwidth while Wi-Fi adopts 20~MHz bandwidth, BLE cannot directly measure Wi-Fi's frequency shift as a BLE channel is always overwhelmingly fulfilled by Wi-Fi signals regardless of Wi-Fi's frequency shifts. In addition, BLE is a single-carrier system that employs a completely different PHY compared to the Wi-Fi PHY. BLE and Wi-Fi cannot directly measure each other's frequency offset using CFO estimators. Considering all these difficulties, we select one level of frequency shift at the opposite directions to enable 1 bit in one packet. Particularly, Wi-Fi packets are shifted with $\pm 100 (130)$~KHz (asymmetric shifts are adjusted according to the overlapping channel) while BLE packets are shifted with $\pm 80$~KHz. Such shift ensures enough space for artificial Doppler shift demodulation while ensuring CFO recovery in legacy packet reception even in the presence of inherent CFO and Doppler effect.

The FCC rules enforce BLE to conduct FHSS by transmitting packets in each of the 40 channels with equal probability. Since the aggregated bandwidth of BLE is several times wider than Wi-Fi channels, it is a certain probability that a BLE packet jumps outside a Wi-Fi channel. To tackle this issue, DopplerFi employs the DSK encoder to guarantee bits are continuously modulated on the overlapped frequencies. Before packet transmission, a selection box in BLE PHY is employed to select the transmission channel according to a pseudo random sequence. The DSK encoder reads channel index chosen by the selection box and assigns artificial Doppler bits to packets on the channels that overlap with the target Wi-Fi channel while skipping packets on other channels to ensure all bits can be captured by the Wi-Fi receiver. 

As illustrated in Fig.~\ref{fig:dsk}, a BLE sender wants to send ``10'' to a Wi-Fi receiver. Meanwhile, it needs to transmit to a BLE receiver. In the first time slot, the selection box chooses channel 7, which is an overlapping channel with Wi-Fi. Thus, the DSK encoder injects an artificial Doppler shift of 80~KHz to the carrier frequency. In the second time slot, BLE sends a packet on a non-overlapping channel (Channel 9), so the DSK buffers the second bit ``0''. In the thirds time slot, the DSK detects that the packet is transmitted on an overlapping channel (Channel 0), it reads the buffer and pick the oldest bit as the artificial Doppler bit to shift the carrier frequency.

\subsection{Extracting Artificial Doppler Shifts From GFSK Demodulator}
\begin{figure}[t]
	\center
	\includegraphics[width=0.5\textwidth]{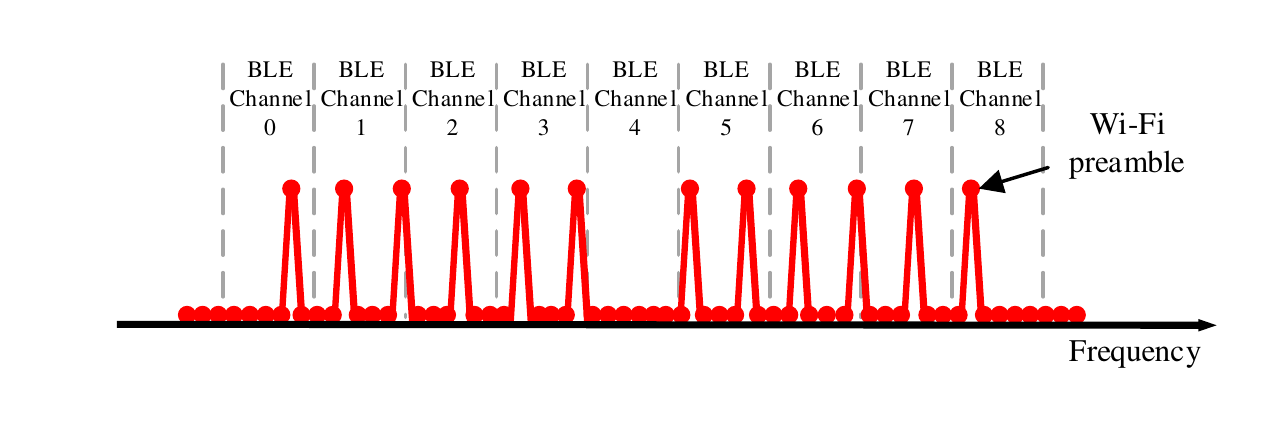}\vspace{0.2cm}
	\caption{Different patterns of Wi-Fi STF received by BLE radios in different overlapping channels.}
	\label{fig:preamble}\vspace{0.3cm}
\end{figure}

Here we introduce a mechanism that allows the BLE receiver to recover the artificial Doppler bits embedded in Wi-Fi packets. To estimate the frequency shift in an incompatible packet sent in a 10$\times$ wider channel, we entail the following hurdles: i) BLE devices employ the GFSK demodulation that cannot be used to decode OFDM symbols in Wi-Fi, and ii) there is no built-in frequency offset estimator that can detect the frequency shift in Wi-Fi packets.

To overcome the above hurdles, our insight is that the preamble prepending to each Wi-Fi packet has distinct structures, whose frequency changes can be reflected in the standard GFSK demodulator in BLE. The short training field (STF) in Wi-Fi preamble consists of 12 subcarriers with unit magnitude and the interval between adjacent non-zero subcarriers is 4~\cite{ieee2010ieee}. Without loss of generality, we use Channel 1 (with central frequency at 2412~MHz) in IEEE 802.11 standard to illustrate. Fig.~\ref{fig:preamble} illustrates the STF frequency patterns received by BLE radios in 9 overlapping channels. As the subcarrier spacing in Wi-Fi is 312.5~KHz and the channel bandwidth in BLE is 2MHz, 6 or 7 subcarriers fall within one BLE channel. BLE receivers in different channels observe different frequency patterns of the Wi-Fi STF. For example, the BLE channel 3~(frequency at 2410MHz) and channel 5 overlap with two non-zero subcarriers while channel 4 overlaps with none. In the DopplerFi system, we select BLE channels that overlap with non-zero subcarriers to receive artificial Doppler bits.

\begin{figure}
	\centering
	\begin{minipage}[b]{0.24\textwidth}\centering
		\includegraphics[width=1\textwidth]{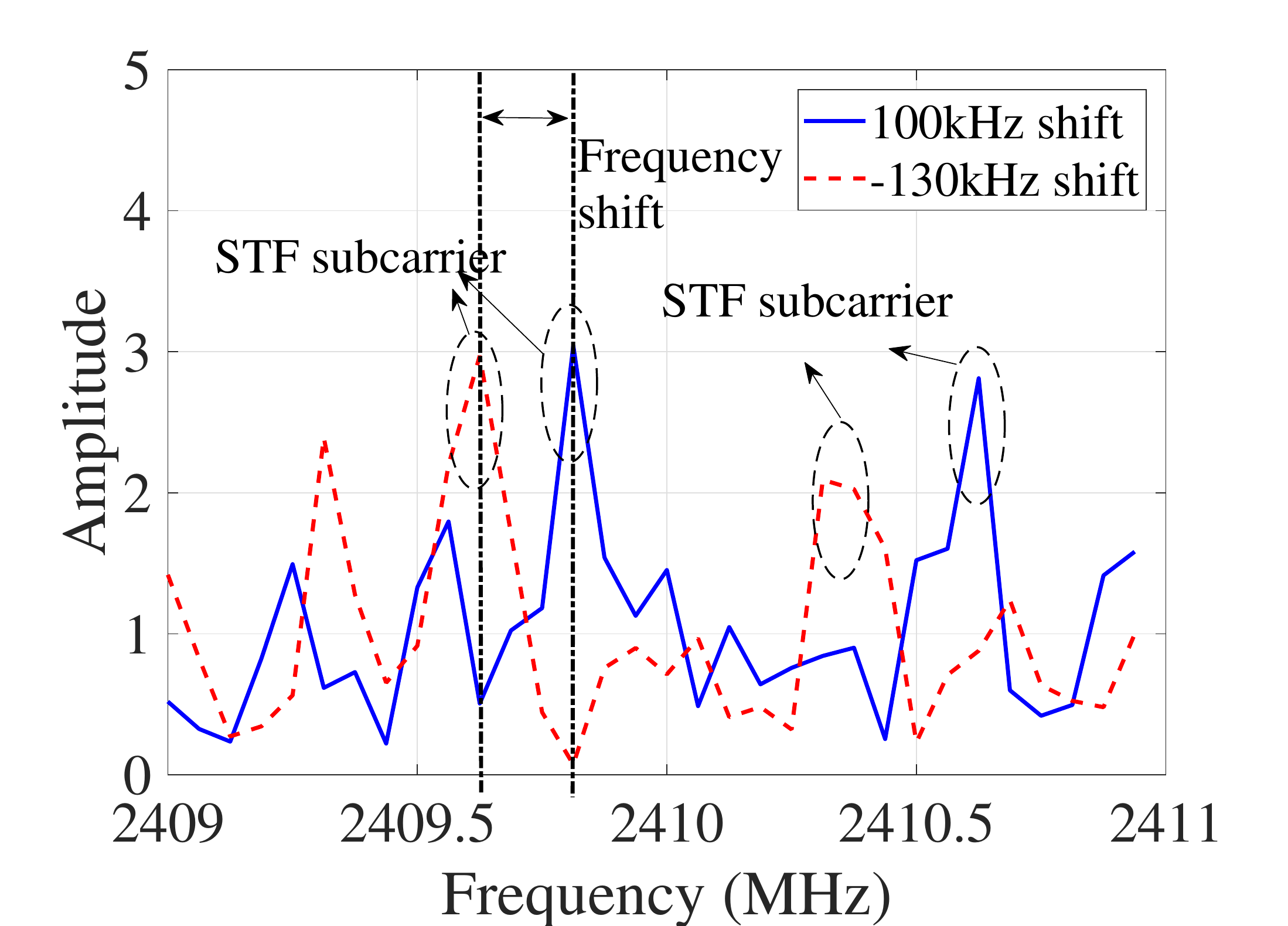}
		\caption{Wi-Fi preamble data captured by a BLE radio. The BLE radio stays in BLE's Channel 3 to receive Wi-Fi packets by a node in Wi-Fi's Channel 1 in the 2.4~GHz band.} \label{fig:freq_pre}
	\end{minipage}
	\begin{minipage}[b]{0.24\textwidth}\centering
		\centering
		\subfigure
		{\label{fig:time_pre1}\includegraphics[width=1\textwidth]{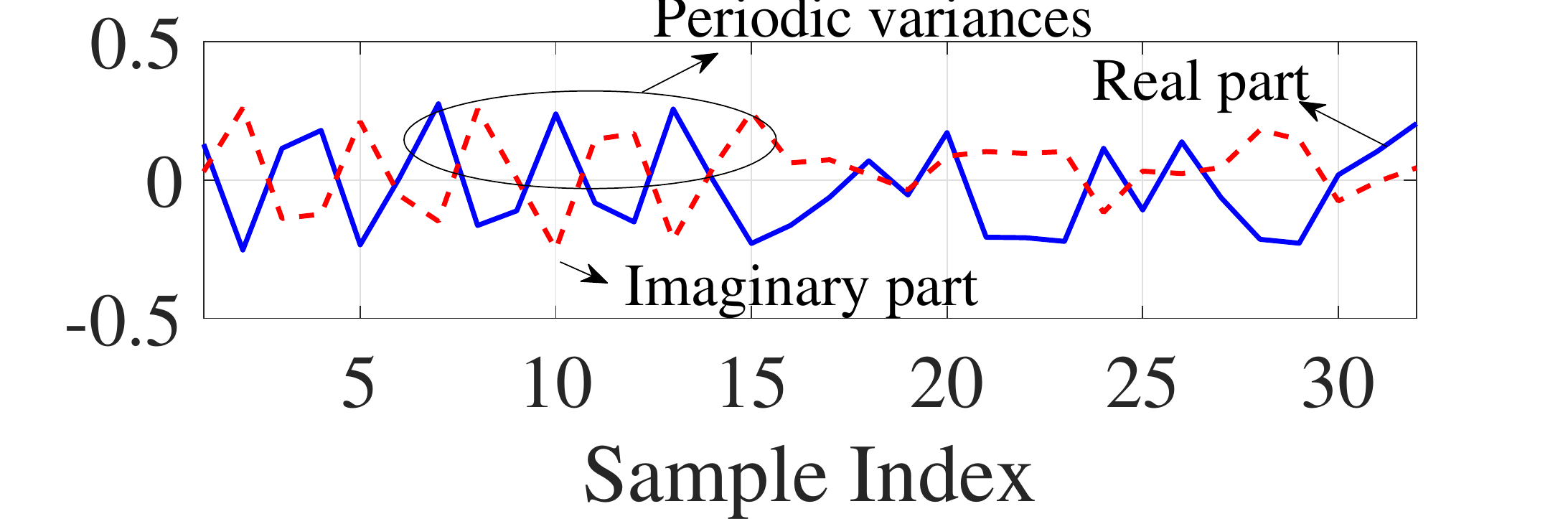}\vspace{0.3cm}}
		\subfigure
		{\label{fig:time_pre2}\includegraphics[width=1\textwidth]{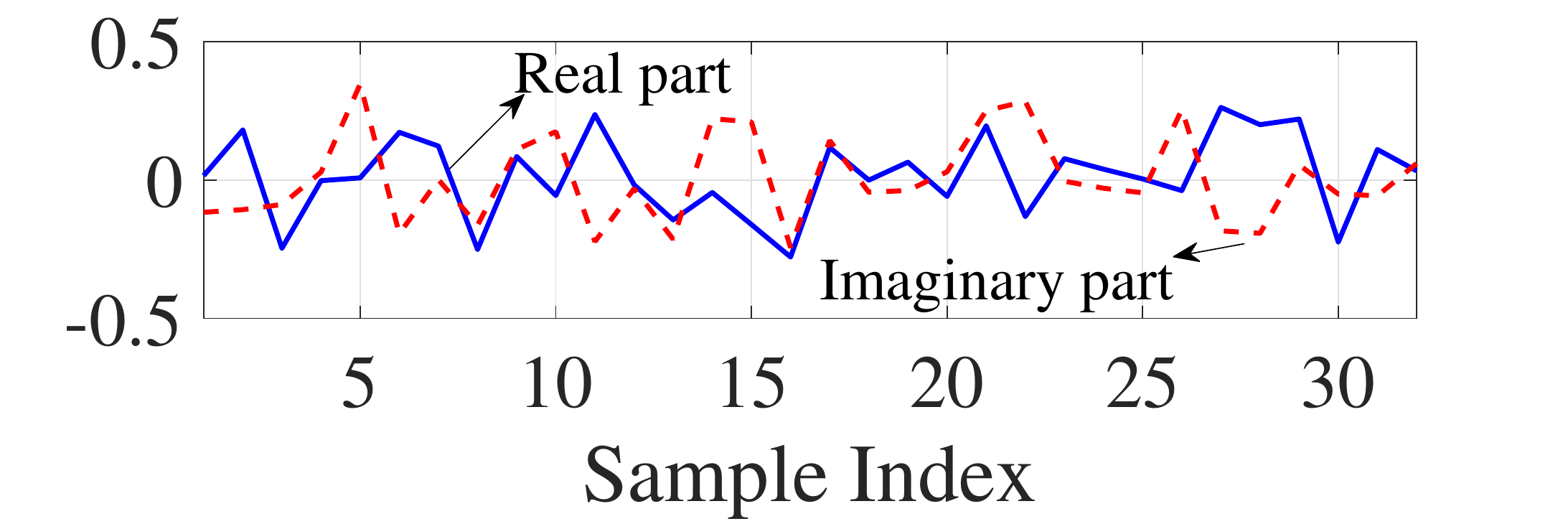}\vspace{0.3cm}}
		\caption{I/Q data of Wi-Fi packets obtained by a BLE radio. Carrier frequency of Wi-Fi packets are shifted by -130~KHz and 100~KHz in the figures from top to bottom.}
		\label{fig:time_pre}
	\end{minipage}
\end{figure}

The above STF patterns enable BLE receivers to differentiate different artificial Doppler shifts in Wi-Fi packets. Fig.~\ref{fig:freq_pre} illustrates two Wi-Fi preambles with different artificial Doppler shifts sampled by a BLE radio in the frequency domain. Although the BLE radio samples at merely 1/10 clock rate compared to the Wi-Fi radio, the captured samples can still reveal the frequency shifts in the non-zero subcarriers. Such frequency shifts are also reflected in the time domain. As depicted in Fig.~\ref{fig:time_pre}, though seemingly irregular, the Wi-Fi preamble samples captured by the BLE radio show distinct periodic patterns in the time domain. 

\begin{figure}[t]
	\center
	\includegraphics[width=0.5\textwidth]{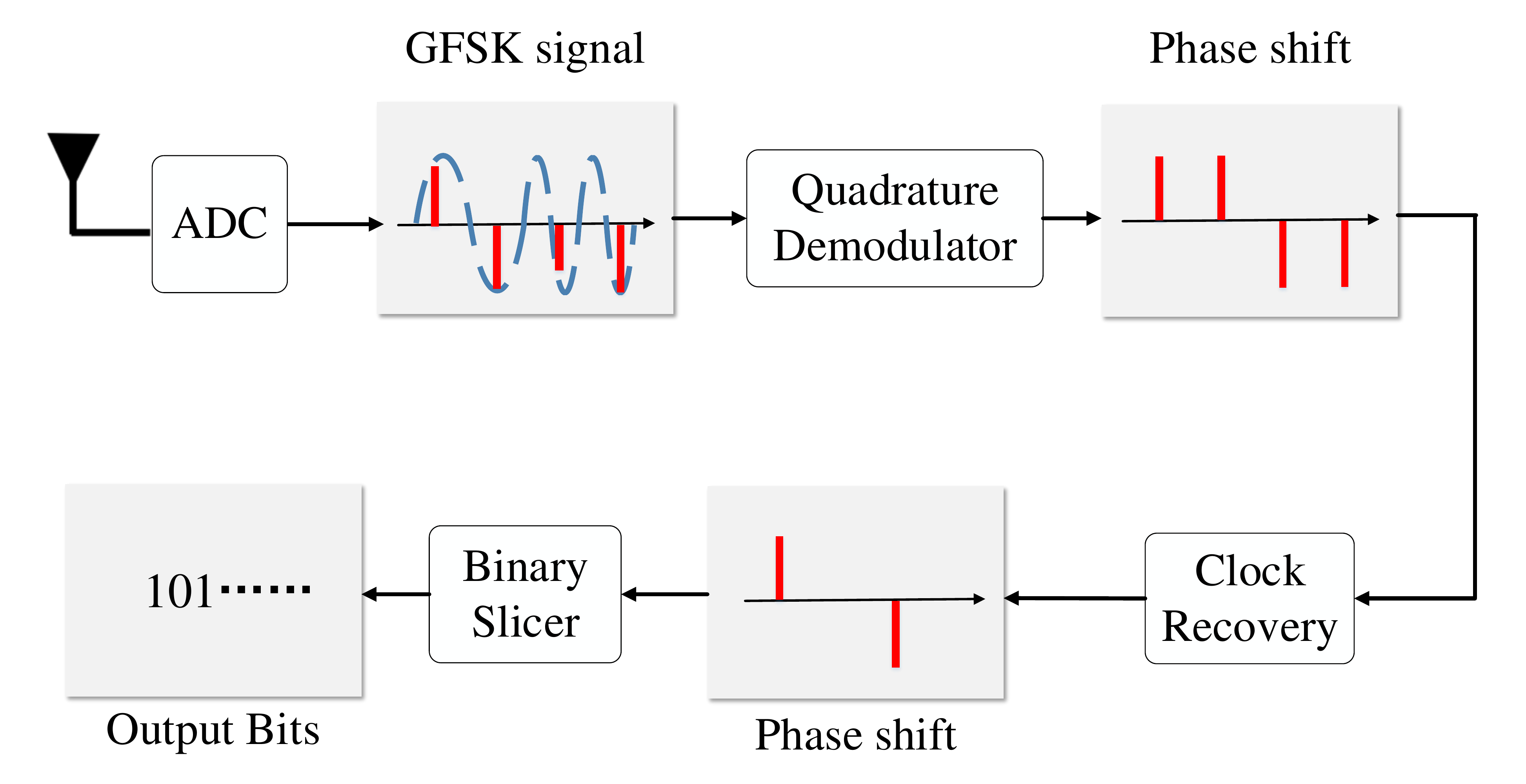}\vspace{0.2cm}
	\caption{BLE GFSK demodulator.}
	\label{fig:GFSK}\vspace{0.3cm}
\end{figure}

Our goal is to extract the above patterns using the data obtainable from standard BLE modules. In particular, we exploit the GFSK demodulator, which is the core signal processing module in BLE receivers. As illustrated in Fig.~\ref{fig:GFSK}, the GFSK demodulator is comprised of three components: quadrature demodulator, clock recovery and binary slicer. The quadrature demodulator is used to demodulate FM, FSK, GFSK signals. Mathematically, it computes the argument of product of the delayed signal and the conjugate undelayed signal as
\begin{equation}
\phi[n] = \arg(x[n]\bar{x}[n-1]), n = 0, 1, ...,
\end{equation}
where $x[n]$ is the baseband signal captured by the receiver and $\bar{x}[n]$ denotes the conjugate signal.
Normally, a GFSK signal $x[n]$ can be written as
\begin{equation}
	x[n] = \exp(2\pi f/f_s n), 
\end{equation}
where $f$ is the frequency of the signal and $f_s$ is the sampling rate of the receiver. Thus, we have $\phi[n]=2 \pi f/f_s$, which reflects the frequency of the signal. Then, $\phi[n]$ is sent to the clock recovery block, which is a discrete-time error-tracking synchronizer that corrects the timing error of the incoming signal. Finally, the corrected phases are put into the binary slicer, where a positive phase is demodulated as ``1'' and a negative phase is demodulated as ``0''.

\begin{figure}
	\centering
	\subfigure
	{\label{fig:output1}\includegraphics[width=0.5\textwidth]{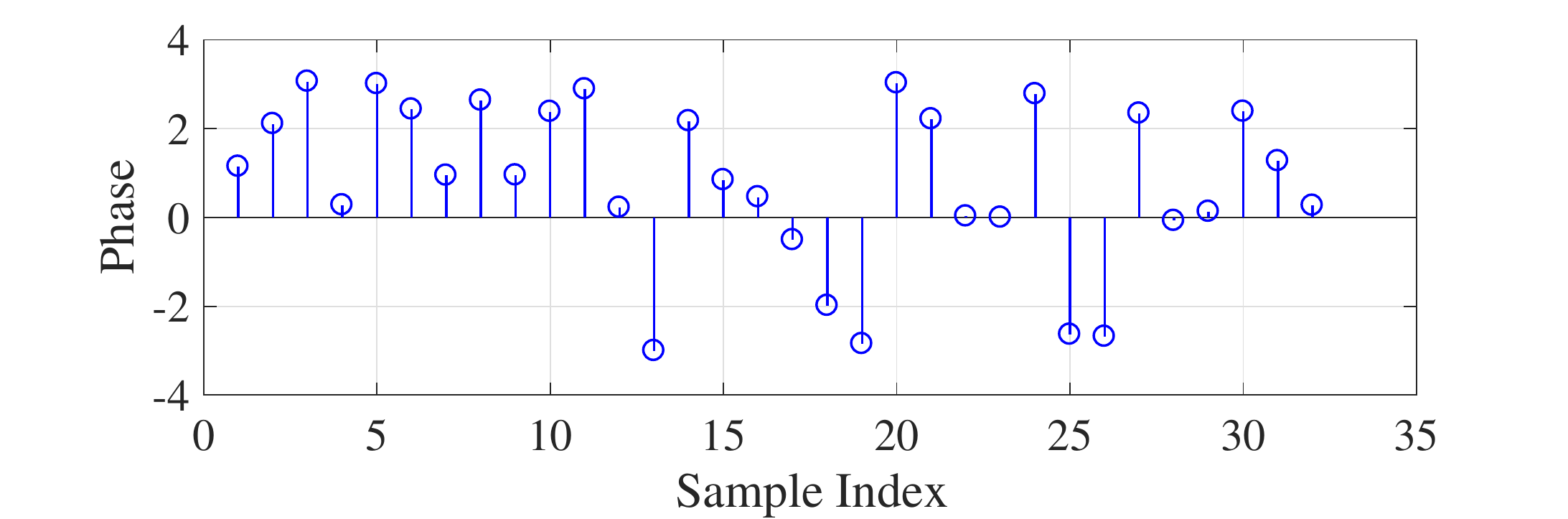}}\vspace{0.3cm}
	\subfigure
	{\label{fig:output2}\includegraphics[width=0.5\textwidth]{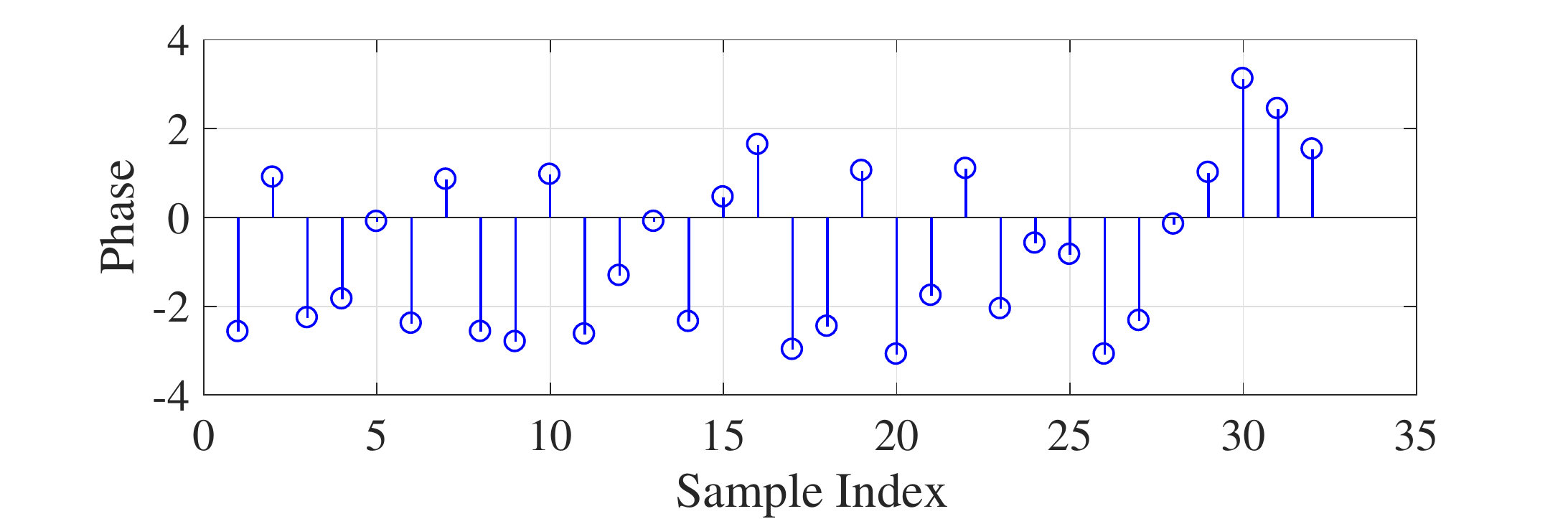}}\vspace{0.3cm}
	\caption{BLE quadrature demodulator output. The BLE radio captures Wi-Fi preamble symbols with -130~KHz and 100~KHz shifts in the figures from top to bottom.}
	\label{fig:output}
\end{figure}

Fig.~\ref{fig:output} shows the output of the quadrature demodulator when the BLE receiver samples Wi-Fi preamble symbols with different artificial Doppler shifts. We observe that frequency shifts in Wi-Fi lead to bias in the quadrature demodulator: positive (negative) phase dominates the output corresponding to the negative (positive) frequency shift. Hence, we can also observe biases in the final output bits, which are obtainable in BLE radios.

To extract the output bits that correspond to Wi-Fi preamble symbols, DopplerFi takes the following two steps.

\textbf{Step 1: GFSK extraction.} First, DopplerFi employs a GFSK extractor to obtain the output bits that correspond to a Wi-Fi preamble. Once a BLE node associate with a Wi-Fi node and is in the receiver mode, it checks the quantized received signal strength indication (RSSI) samples to detect the start of a Wi-Fi packet. Let $RSSI[n]$ represent the received RSSI value of the $n$th sample. If the RSSI is lower than a threshold while the delayed RSSI is higher than the threshold, we regard that the start of a Wi-Fi packet is successfully detected. In particular, the GFSK extractor continuously checks the following condition
\begin{align}
{RSSI}[n-1] < {RSSI}_\text{Threshold} < {RSSI}[n].
\end{align}
The threshold $\text{RSSI}_\text{Threshold}$ is empirically set to 0.02 in our experiment.
After we detect the start of a Wi-Fi frame, we stack the first 16 bits output by the binary slicer. 

\textbf{Step 2: GFSK demapping.} Next, DopplerFi determines the artificial Doppler bit from the bit sequence obtained by the GFSK extractor. The rule is to check the bias lying in the bit sequence, as described below.
\begin{equation}
\begin{split}
\begin{cases}
\sum_{n=0}^{n=15}{o[n]}> \eta & \Rightarrow 0 \\
\sum_{n=0}^{n=15}{o[n]}\leq \eta & \Rightarrow 1
\end{cases},
\end{split}
\label{e:central_limit}
\end{equation}
where the $o[n]$ is the output sample of the BLE GFSK demodulator. We sum the first 16 samples of the output because the length of the received Wi-Fi preamble is 16 samples.
The decision gate $\eta$ is related to the transmission channel. We empirically set the value to eight in our experiments, which optimizes the decoding accuracy.

\subsection{Extracting Artificial Doppler Shifts From CSI}

\begin{figure}
	\centering
	\subfigure
	{\label{fig:spectrum}\includegraphics[width=0.5\textwidth]{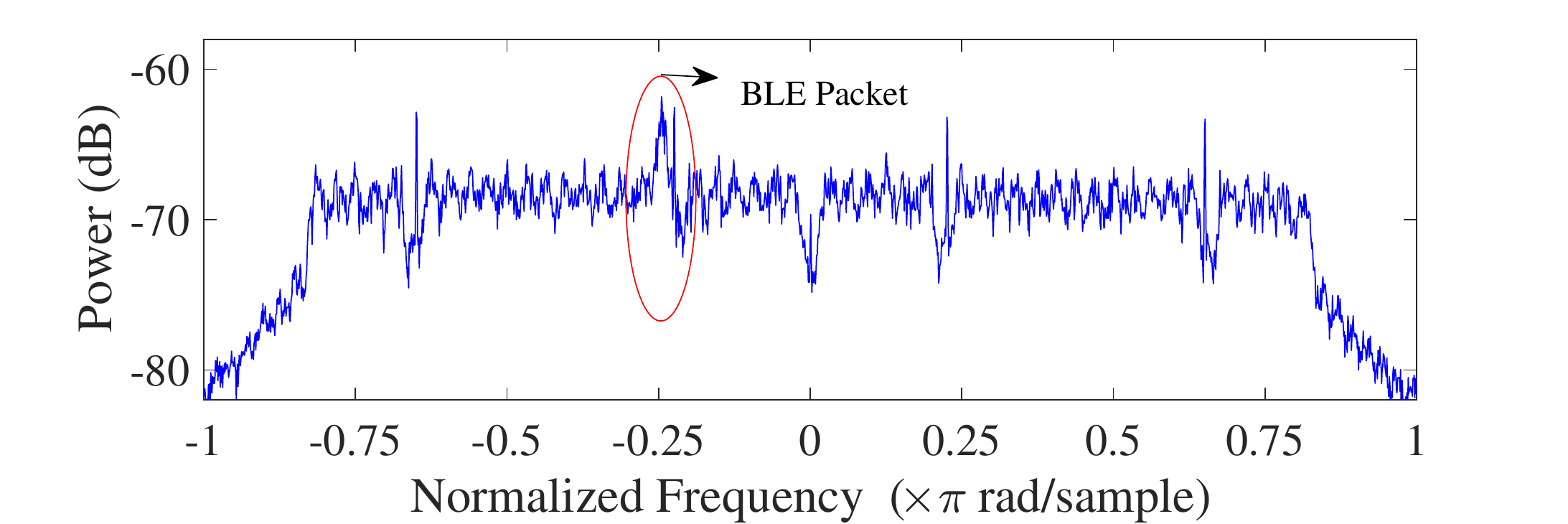}}
	\hspace{0.2cm}
	\subfigure
	{\label{fig:hsy}\includegraphics[width=0.51\textwidth]{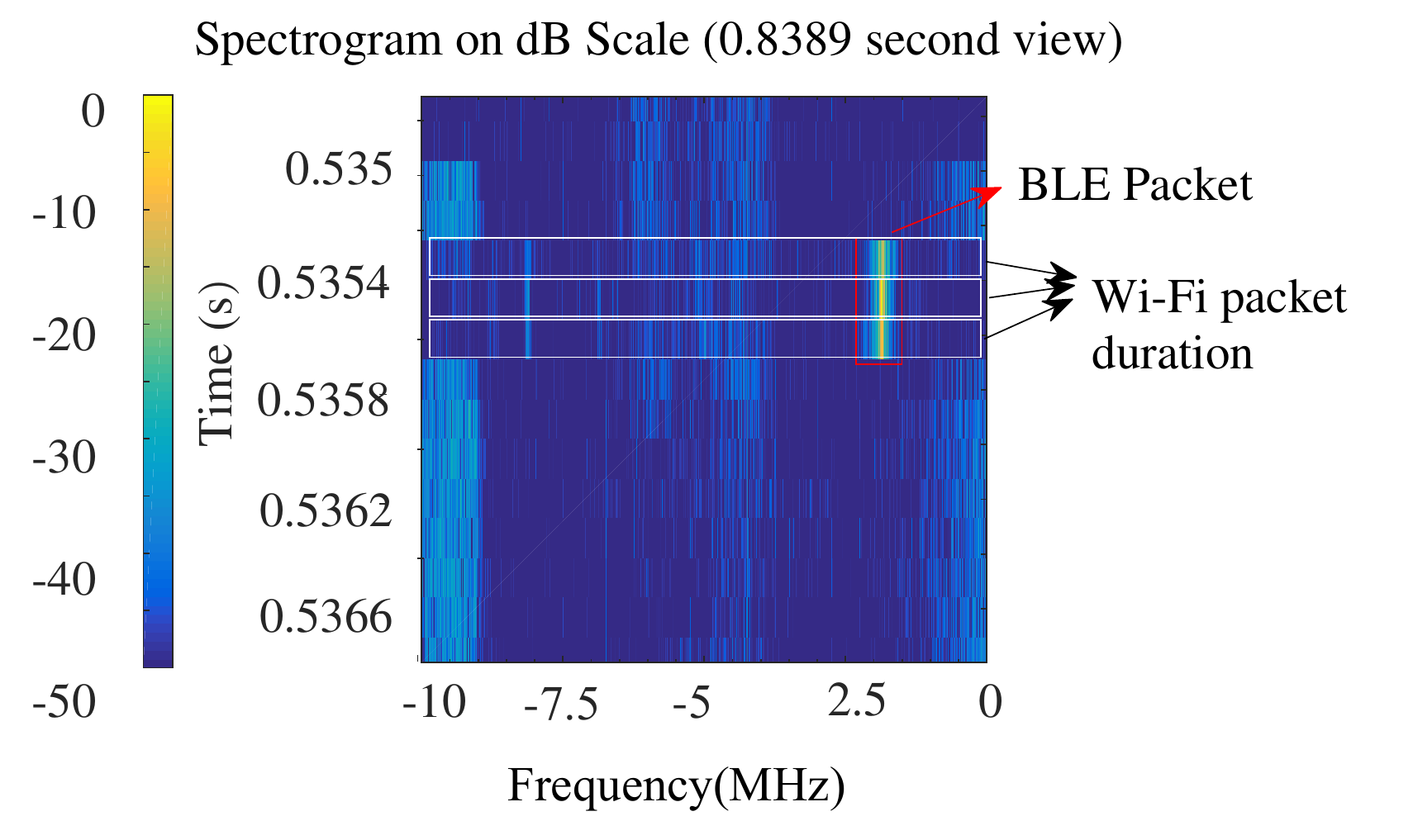}\vspace{-0.1cm}}
	\caption{Spectrum of a Wi-Fi channel captured by WARP. A BLE packet hits overlapped subcarriers of multiple Wi-Fi packets.}
	\vspace{0.3cm}
	\label{fig:wifibleoverlap}
\end{figure}

Now we present how a Wi-Fi receiver demodulates the artificial Doppler bits in BLE packets. Since the BLE packets is transmitted in much narrower channels that overlap with only several subcarriers, we utilize CSI, which is accessible from commercial Wi-Fi cards, to estimate frequency shifts in BLE packets. To this end, we have the following observations. First, CSI values in several Wi-Fi subcarriers will be affected by BLE packets, resulting in energy spikes or deep drops, as illustrated in the top figure of Fig.~\ref{fig:wifibleoverlap}. Second, the time duration of a BLE packet is usually longer than the duration of a Wi-Fi packet, and thus affects CSI values of multiple Wi-Fi packets. The spectrogram in Fig.~\ref{fig:wifibleoverlap} shows that a BLE packet overlaps with three consecutive Wi-Fi packets.

Based on these observations, DopplerFi demodulates artificial Doppler shifts in BLE packets using CSI as follows.

\textbf{Step 1: CSI extraction.} DopplerFi first employs the CSI Extractor to identify the CSI values that are hit by BLE packets. As adjacent subcarriers experience very similar multipaths and have similar wavelengths, the CSI values across adjacent subcarriers vary smoothly except for BLE affected ones. Therefore, we locate the BLE affected CSI values with the following equation:
\begin{equation}
	D[k] =  |{CSI}[k] - {CSI}[k-1]|,
\end{equation}
where $CSI[k]$ indicates the CSI value on the $k$th subcarrier. If the maximum value in $D[k]$ is higher than a threshold, the CSI Extractor identifies the corresponding subcarriers as the ones hit by a BLE packet and logs the CSI values. 

\textbf{Step 2: CSI Demapping.} Then, DopplerFi demodulates the artificial DopplerFi bits according to the logged CSI values using the CSI demapper module. Normally, there are multiple CSI vectors corresponding to one BLE packet. Therefore, after fetching the subset of the extracted CSI values, we first pick up the amplitude for each CSI vector and then locate the subcarrier with the maximum difference as follows.
\begin{equation}
	I = \arg \max_{k}|{CSI}[k] - \overline{{CSI}}|,
\end{equation}
where $\overline{\text{CSI}}$ is the mean of CSI values on non-zero subcarriers. If there are multiple CSI vectors within one BLE's transmission time slot, we average all the indexes $\{I\}$. Analogous to the GFSK demapper, the index or averaged index is compared with the standard BLE frequency to determine the bias caused by artificial Doppler shifts. Bias to lower (higher) frequency subcarriers is interpreted as ``0'' (``1'').

\subsection{Synchronization}

Thus far, we have elaborated the encoding and decoding processes using the Artificial Doppler shifts. In this section, we design a synchronization method based on the free side channel. 

The association or pairing mechanisms in BLE and Wi-Fi are incompatible and cannot be directly applied to DopplerFi to initiate a cross-technology channel between a Wi-Fi node and a BLE node. A BLE node cannot simply scan all channels to identify the channel adopted by the target Wi-Fi node, as the BLE node cannot decode Wi-Fi beacons or data packets. To tackle this issue, we allow the BLE node to concurrently send one probe request to each Wi-Fi channel while following the mandatory FHSS sequences. Once the DopplerFi-enabled Wi-Fi node receives the probe request, it sends back a probe response to the BLE node. Then, the association between the BLE and Wi-Fi nodes is established.

In the DopplerFi system, the sender generates a lightweight preamble when it starts transmission. Particularly, the preamble is designed as a fixed sequence of alternating 0 and 1. We use the alternating 0 and 1 bits to represent the preamble because it is different from the Doppler effect caused by the crystal oscillator and device mobility. The preamble is designed as 4 bits to trade off the throughput and false detection probability. For each DopplerFi transmission, the sender prepends a preamble before payload bits. The DopplerFi receiver first senses the energy of the channel and quantizes the RSSI values to indicate whether it is a DopplerFi bit or idle channel. Next, the DopplerFi node tests the followed 4 DopplerFi bits to judge whether the preamble is detected or not. 

\subsection{Integration with Legacy Networks}
Thus far we have elaborated all design components of DopplerFi. Now we discuss some practical issues when integrating DopplerFi with legacy Wi-Fi and BLE networks. 

\textbf{DopplerFi MAC.} DopplerFi nodes only piggyback artificial Doppler bits when they have legacy traffic. Thus, the MAC behaviors of DopplerFi completely conforms to channel access of legacy Wi-Fi and BLE nodes. When legacy nodes do not have enough data traffic, DopplerFi utilizes mandatory control frames such as beacons to piggyback artificial Doppler bits.

\textbf{Inherent CFO and Doppler effect.} The inherent CFO is caused by impairments in oscillators and varies over time. The CFO is within a limited range of 400~Hz~\cite{fica}, which is much smaller than the artificial Doppler shift. Recent developments in elements and circuits make CFO even smaller. Experiments show that this amount of CFO cannot be recognized by DopplerFi and does not affect the demodulation performance. 
Doppler effects caused by indoor movements are merely several Hertz or tens of Hertz, and thus are negligible to DopplerFi.

\textbf{Impact on adjacent channel.} Adjacent Wi-Fi channels are separated by an interval termed as guard band, which is a 2~MHz gap to avoid adjacent channel interference. DopplerFi induces frequency shifts that less than 5\% of the guard band width. Such an amount of frequency shifts have been considered in the Wi-Fi design and cause no or negligible cross-channel interference. BLE transmits data using a single carrier with $\pm$250~KHz shifts in one channel. Since the BLE channel spacing is 2~MHz, the gap between two adjacent BLE channels is 1.5~MHz (taking into account the frequency shifts caused by the GFSK modulation), which can tolerate ten times of the frequency shifts induced by DopplerFi.

\textbf{Transmission range.} The transmission range is related to the output power of transceiver. DopplerFi applies to a potential solution of IoT devices with BLE protocol. Specifically, we used a Bluetooth Adapter to send BLE packet with DopplerFi bits. The advantage of Ubertooth is the open source of the signal process in MAC Layer so that Artificial Doppler Shift can be embedded as we needed. However, in the test environment, the output power set in the transceiver of Ubertooth is -7dBm, which is significantly lower than the maximum power(4dBm) of the commercial BLE chip. A commercial BLE device can be more long communication distance so that DopplerFi system can be deployed in the indoor environment.

\section{Evaluation}
\label{sec:evaluate}

\subsection{Implementation and Experimental Setup}
\label{sec:implement}

\begin{figure}
	\centering
	\subfigure[DopplerFi prototype based on TI CC2400, USRP and WARP.]
	{\label{fig:device}\includegraphics[width=0.22\textwidth]{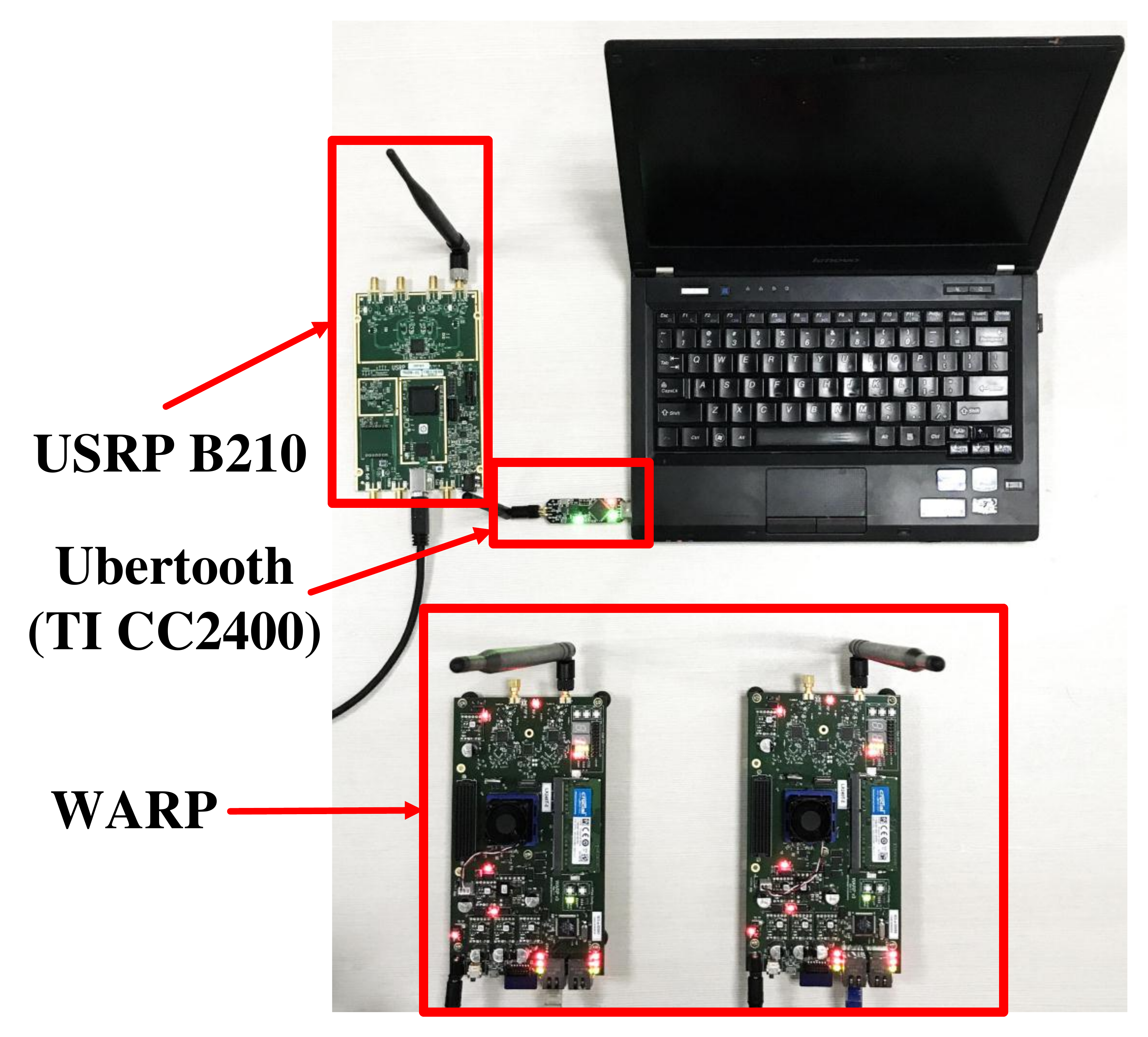}}
	\hspace{0.2cm}
	\subfigure[Testbed topology.]
	{\label{fig:floorplan}\includegraphics[width=0.24\textwidth]{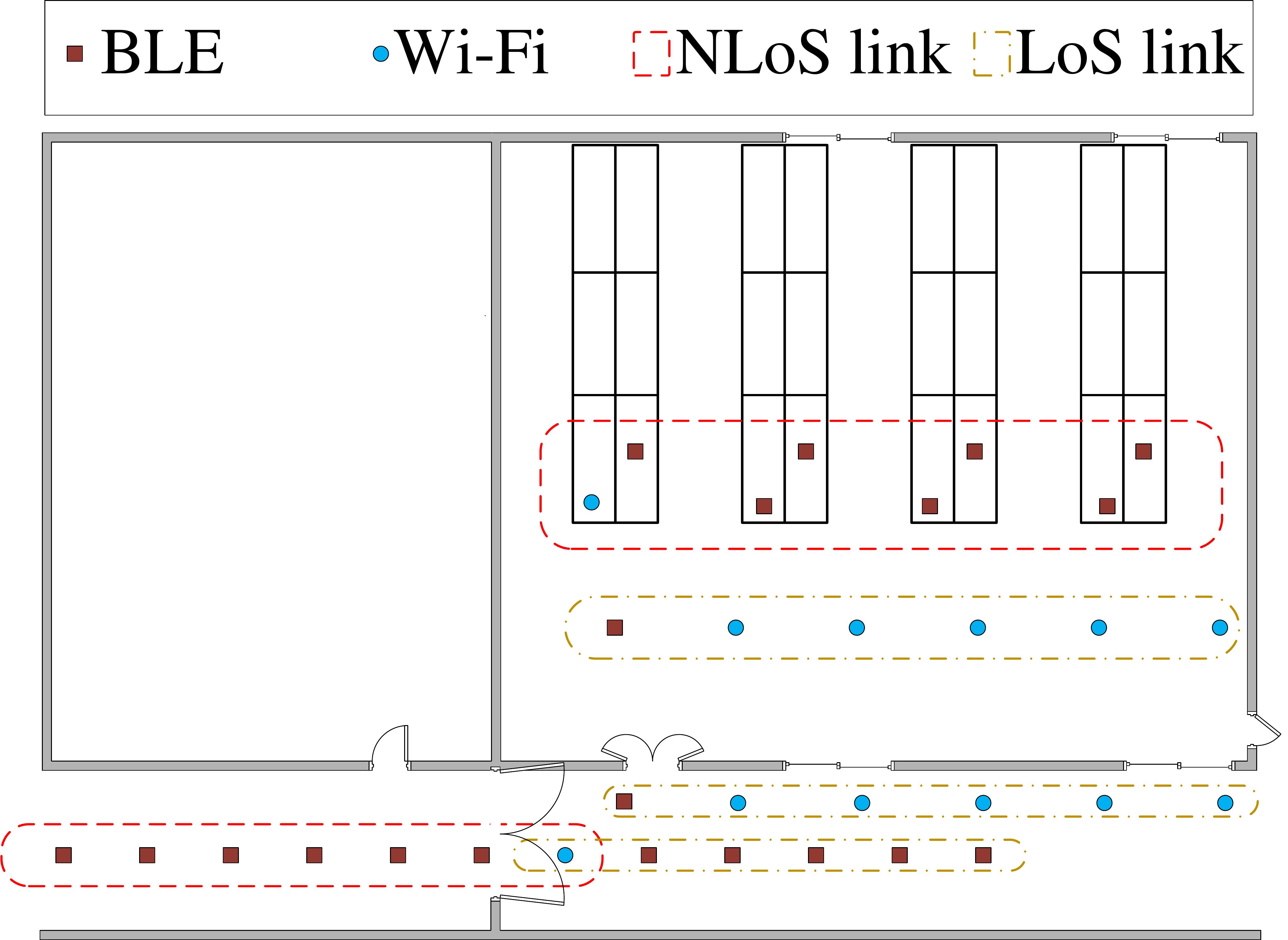}}
	\vspace{0.3cm}
	\caption{Experiment setting and environment.}
\end{figure}
We implement DopplerFi on three platforms: WARP, Ubertooth, and USRP, as shown in Fig.~\ref{fig:device}. WARP nodes run the open source 802.11 reference design that conforms to standard Wi-Fi PHY/MAC. Ubertooth is a BLE adapter equipped with a LPC175x ARM Cortex-M3 MCU operating speed up to 120MHz and a TI CC2400 transceiver in 2.4~GHz~\cite{ubertooth}. USRP nodes are integrated with GFSK-based BLE PHY, which enables precise exploration of intermediate results in BLE PHY. We inject artificial Doppler shifts by dynamically changing the frequency offset in these devices. We modify the MDMCTRL register in TI CC2400 to change the frequency offset of Ubertooth. We modify the 802.11 reference design to change the frequency offset in WARP.

We conduct experiments in a typical indoor office environment as shown in Fig.~\ref{fig:floorplan}. We deploy Wi-Fi and BLE nodes at different locations in an office or a hallway, which are separated by a stone wall. The BLE and Wi-Fi links are tested under both line-of-sight (LOS) and non-line-of-sight (NLOS) scenarios. There are 20+ Wi-Fi APs in proximity. 26 persons sit in the office area, while a few persons were walking during the experiments. The channel access of nodes is configured to conform to standard 802.11 MAC and BLE link layer. Unless otherwise specified, Wi-Fi nodes are configured to operate on channel 1 centered at 2.412~GHz.

\subsection{Impact on Legacy Transmissions}
The essential motivation of DopplerFi is that there is sufficient redundancy in carrier frequencies in that injecting a controllable amount of frequency shift has a negligible impact on legacy transmissions. This set of experiment evaluates the transparency of DopplerFi. We set different amounts of frequency shift to modulate artificial Doppler bits in Wi-Fi/BLE senders, which send files to the corresponding legacy receivers.

\begin{figure*}
	\centering
	\hspace{-2cm}
	\begin{minipage}[b]{2.2in}
		\includegraphics[width=2.1in]{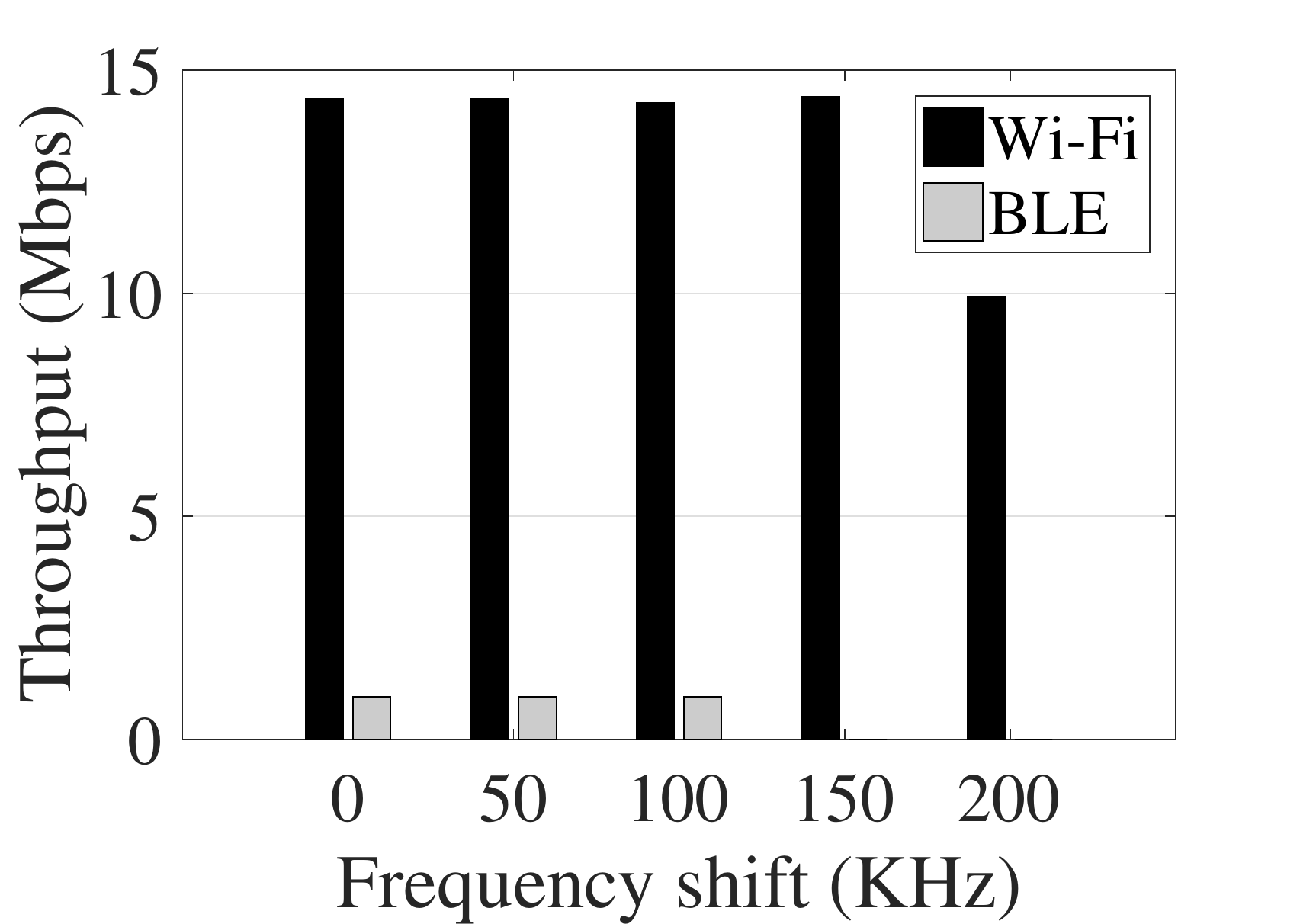}
		\vspace{0.4cm}
		\caption{Throughput of legacy transmissions embedded with DopplerFi.}
		\label{fig:transparant_for_upper}
	\end{minipage}
	\hspace{0.4cm}
	\begin{minipage}[b]{4.4in}
		\subfigure[Wi-Fi to BLE.]
		{\includegraphics[width=2.1in]{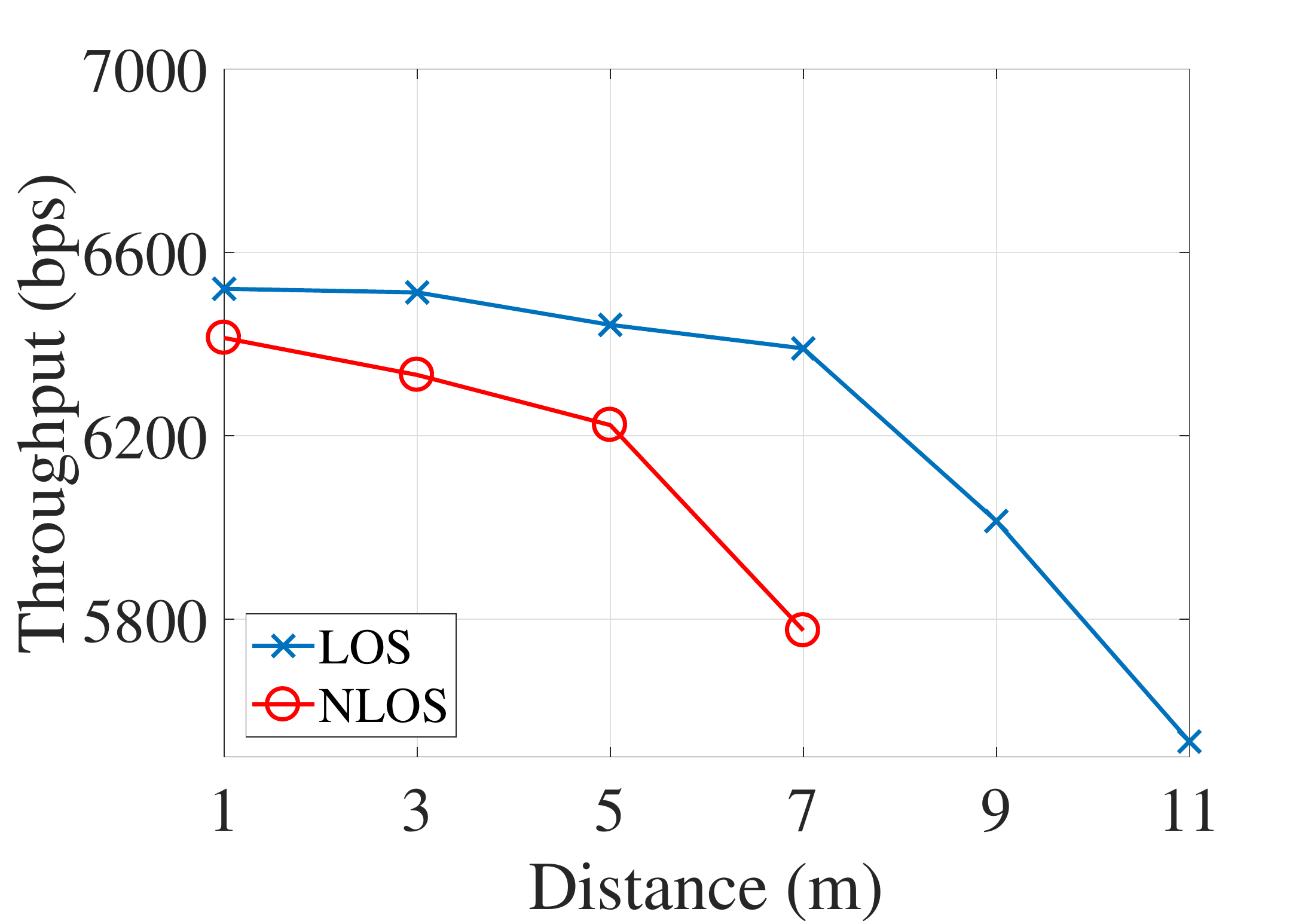}
			\label{throughput_wifi_continuous}}
		\hspace{0.4cm}	
		\subfigure[BLE to Wi-Fi.]
		{\includegraphics[width=2.1in]{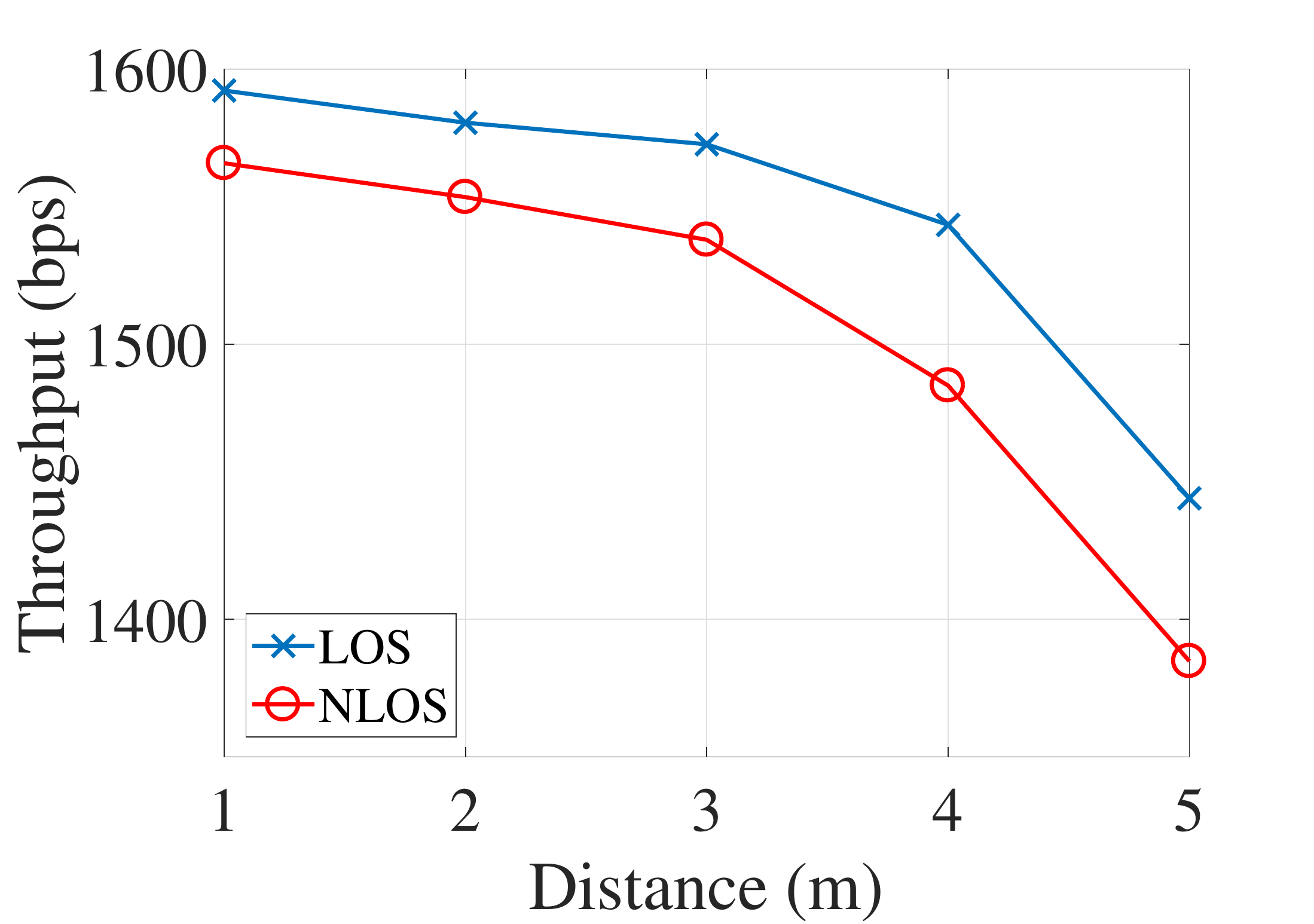}
			\label{throughput_ble_continuous}}
		\caption{Throughput at various distances in the ideal environment.}
		\label{fig:throughput_continuous}
	\end{minipage}
	\hspace{-2cm}
	\vspace{-0.3cm}
\end{figure*}

\begin{figure*}
	\centering
	\begin{minipage}[b]{3.5in}
		\subfigure[Wi-Fi to BLE.]
		{\includegraphics[width=1.65in]{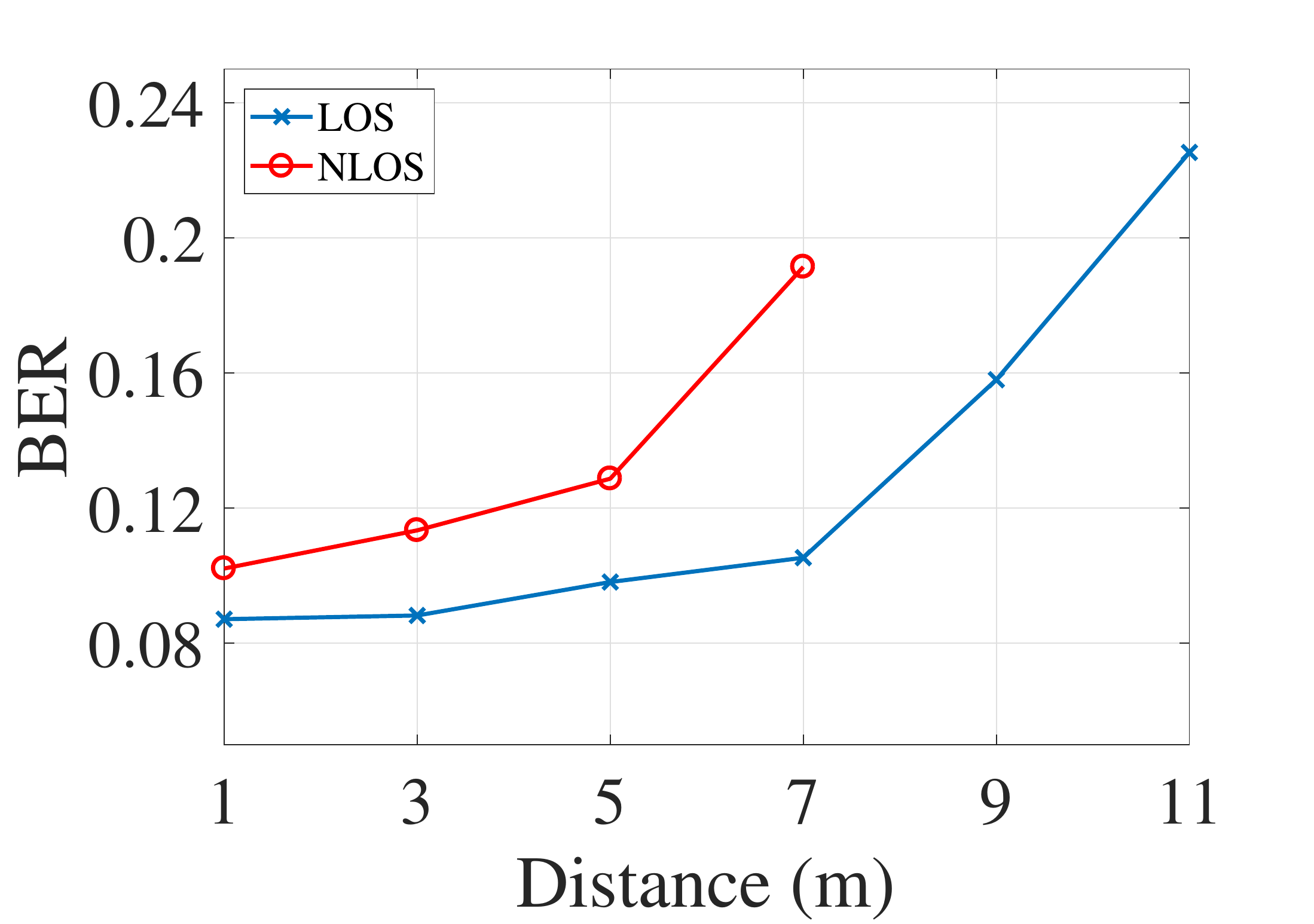}
			\label{wifi_ber}}
		\subfigure[BLE to Wi-Fi.]
		{\includegraphics[width=1.65in]{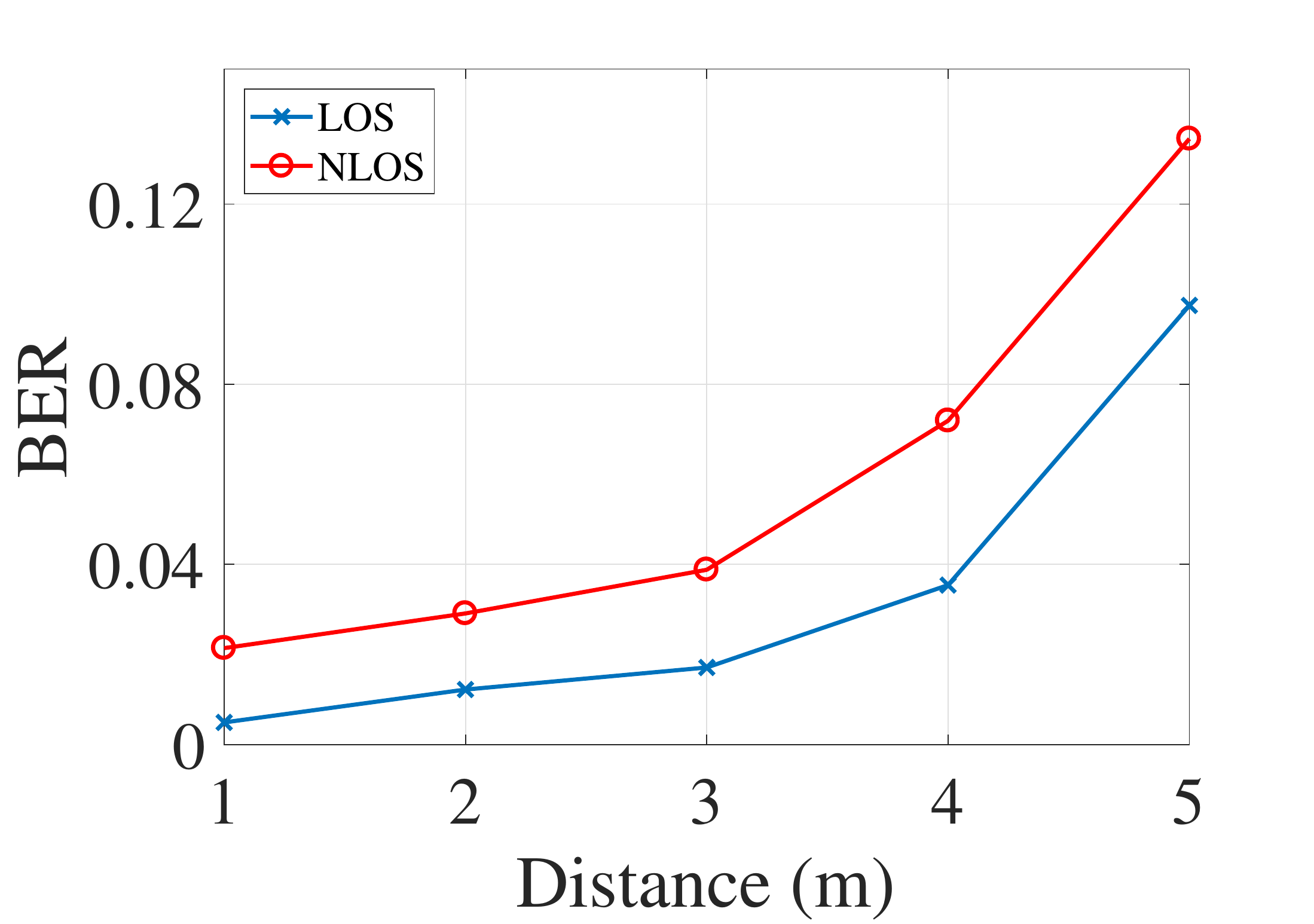}
			\label{ble_ber}}	
		\caption{BER vs. communication distance.}
		\label{fig:BER}
		\vspace{0.2cm}
	\end{minipage}
	\begin{minipage}[b]{3.5in}
		\subfigure[Wi-Fi to BLE.]
		{\includegraphics[width=1.65in]{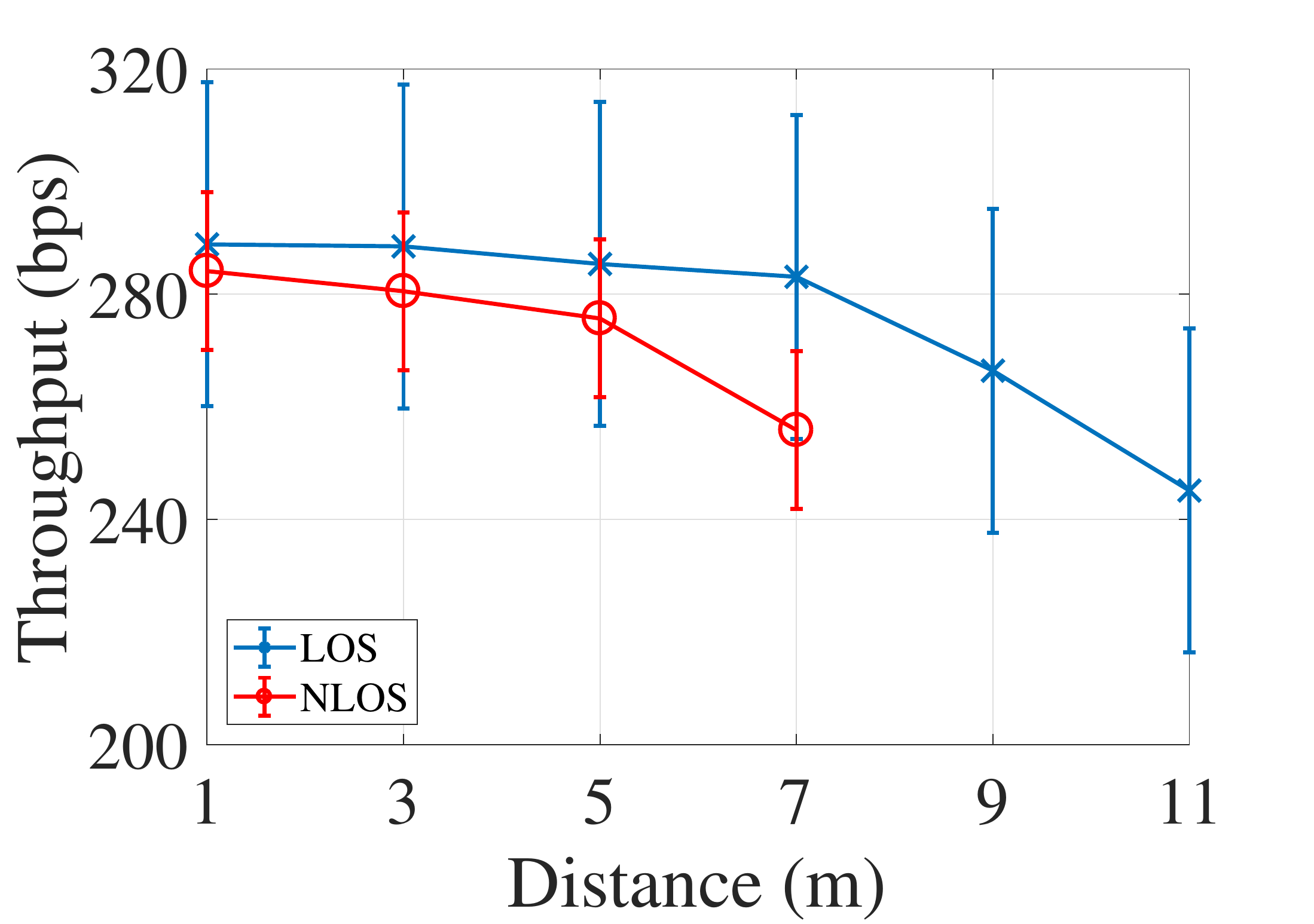}
			\label{throughput_wifi_transparent}}
		\subfigure[BLE to Wi-Fi.]
		{\includegraphics[width=1.65in]{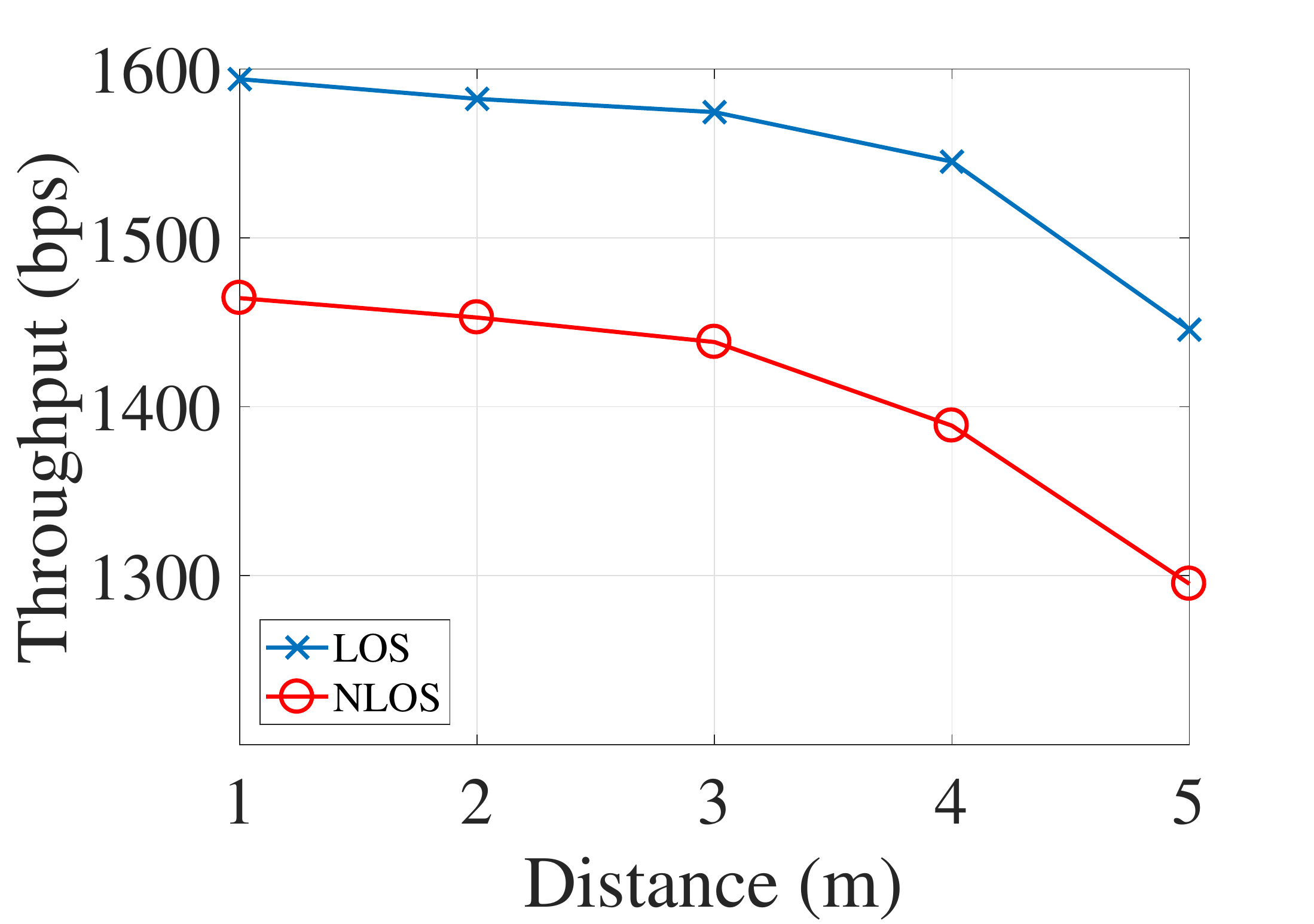}
			\label{throughput_ble_transparent}}
		\caption{Throughput vs. communication distance.}
		\label{fig:throughput}
		\vspace{0.2cm}
	\end{minipage}
\end{figure*}

Fig.~\ref{fig:transparant_for_upper} illustrates the throughputs of legacy Wi-Fi and BLE links embedded with different amounts of artificial Doppler shift. We observe that when we inject no more than 150~KHz frequency shifts, the legacy Wi-Fi link suffer a negligible amount of throughput loss (less than 0.8\%). The legacy BLE link can tolerate up to 100~KHz frequency shift with throughput loss less than 0.3\%. DopplerFi injects 100~KHz and 80~KHz frequency shifts into Wi-Fi and BLE senders, and thus have a negligible impact on legacy transmissions.

\subsection{Interference-Free Environment}
To test the performance limit of DopplerFi, we configure WARP and USRP nodes to continuously send packets in a clean channel with minimal interference. This set of experiments is conducted at midnights, during which there are only few active links in the environments. According to FCC's regulations, the packet interval in BLE is fixed to be 625~$\mu s$. The length of Wi-Fi packet is set to 100~$\mu s$ with packet interval of 40~$\mu s$.

Fig.~\ref{fig:throughput_continuous} shows the throughput of DopplerFi under the ideal environment. We observe that the maximum achievable throughputs for Wi-Fi to BLE (W2B) and BLE to Wi-Fi (B2W) are over 6.5~Kbps and 1.59~Kbps, respectively. For LOS links, W2B and B2W retain at least 6.3~Kbps and 1.54~Kbps throughputs within the transmission ranges of 9~m and 4~m, respectively. The maximum achievable throughput is comparable to the state of the arts~\cite{b2w2,cmorse,freebee}.

\subsection{Real Environment}

We evaluate the performance of DopplerFi in real environments, where there are 20+ APs contending channel with DopplerFi links. The error bar is depicted according to standard deviation. We thoroughly study the BER and throughput of DopplerFi transmissions under a wide range of scenarios.

\textbf{Impact of communication distance.} We first investigate the performance of DopplerFi in both LoS and NLoS scenarios with different communication distances. Fig.~\ref{wifi_ber} shows the BER of W2B links in both LOS and NLOS scenarios. The BER of W2B link in LOS scenarios grows linearly from 8.71\% to 11.99\% when the distance increases from 1~m to 9~m, while the BER increases sharply when the distance reaches 11~m. The BER under NLoS also has a similar trend, while reaching such a critical point at 5~m. The BER in NLoS scenarios is about 1.17$\times$ of the BER in LOS scenarios.

Fig.~\ref{ble_ber} depicts the BER of B2W links at various communication distances. Compared to W2B links, B2W links achieve lower BER at close distances within 4~m while suffering higher BER when the communication distance exceeds 4~m. The reason is that CSI demodulator in Wi-Fi can precisely recover the frequency shifts in BLE only when the magnitude of BLE samples is large enough. The BER gap in B2W between NLOS and LOS scenarios is up to 4.37$\times$, which is larger than that in W2B. This is because the received signal strength drops significantly through wall in NLOS scenarios.

Throughput performance at various distances is shown in Fig.~\ref{fig:throughput}. We observe that throughput in the real environment is much lower compared to the ideal environment without interference. This is because DopplerFi senders need to contend channel with 20+ APs and their associated clients, and thus transmission opportunities are much lower. Compared with B2W, W2B can transmit at a longer distance while yielding lower throughput. This is because BLE is less affected by surrounding Wi-Fi links.

\textbf{Impact of device mobility.} In mobile scenarios, device motion incurs inherent Doppler shifts, which may has impact on the overall frequency shifts at receivers. This set of experiments test the impact of inherent Doppler shifts on the decoding performance. 
We carry the device to walk around in the indoor environments to emulate a person's daily activities. The difference in BER between the mobile and static scenario is less than $2.4\times 10^{-4}$.

\begin{figure*}
	\centering
	\begin{minipage}[b]{3.5in}
		\subfigure[Wi-Fi to BLE.]
		{\includegraphics[width=1.65in]{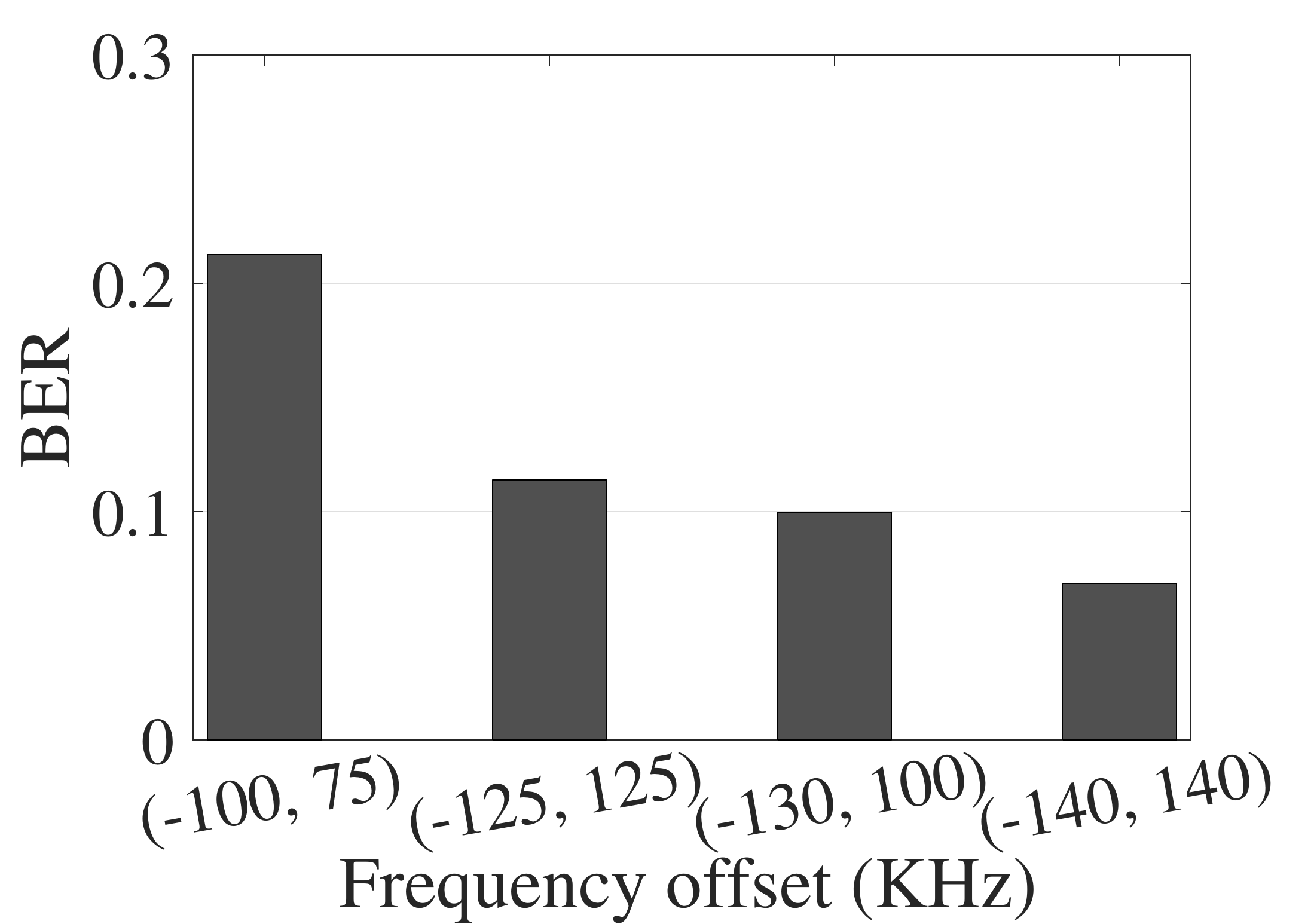}
		\label{ber_wifi_freq}}
		\subfigure[Wi-Fi to BLE.]
		{\includegraphics[width=1.65in]{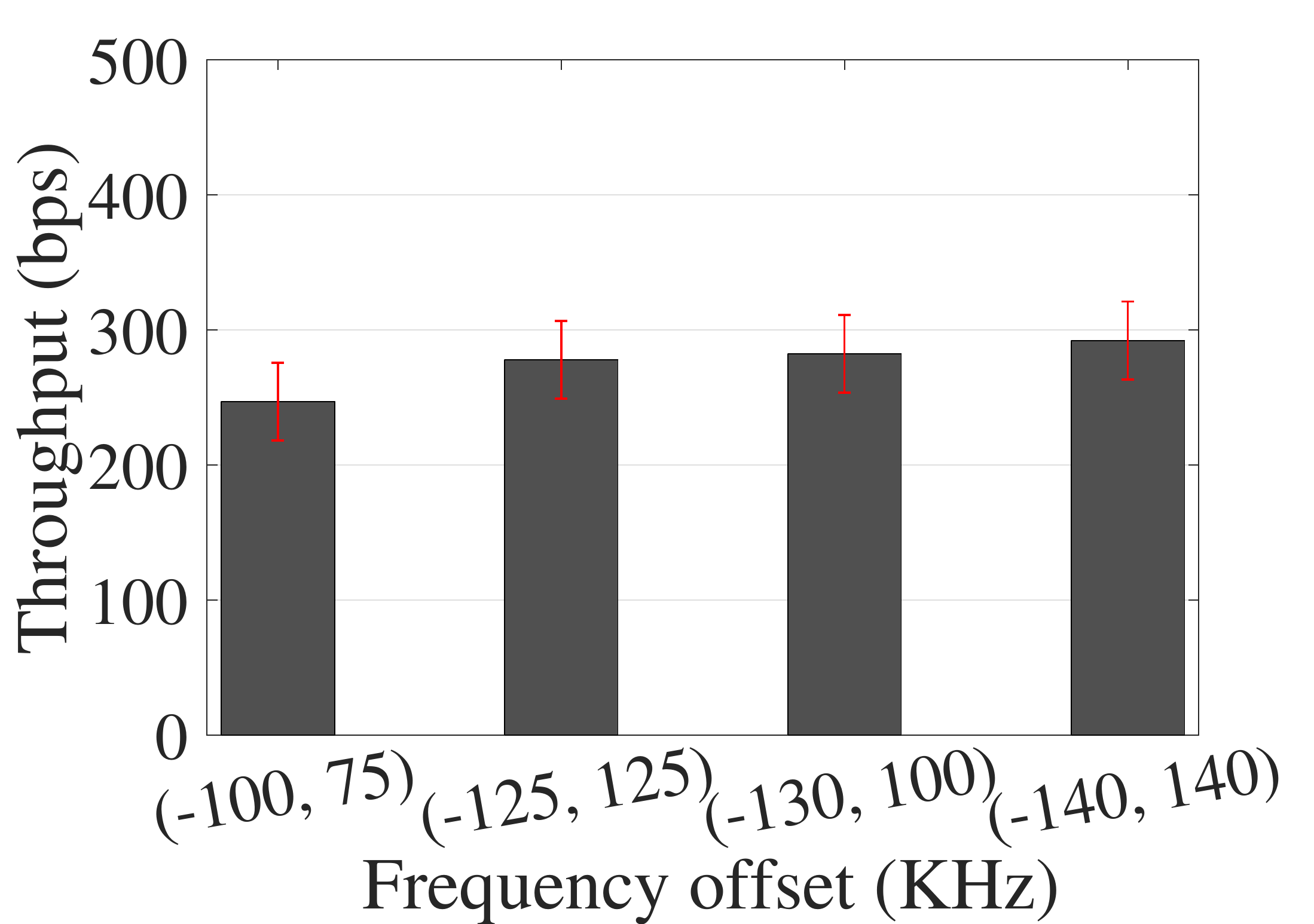}
		\label{th_wifi_freq}}	
		\caption{Impact of frequency shift on Wi-Fi to BLE links.}
		\label{fig:w2b_freq}
	\end{minipage}
	\begin{minipage}[b]{3.5in}
		\subfigure[BLE to Wi-Fi.]
		{\includegraphics[width=1.65in]{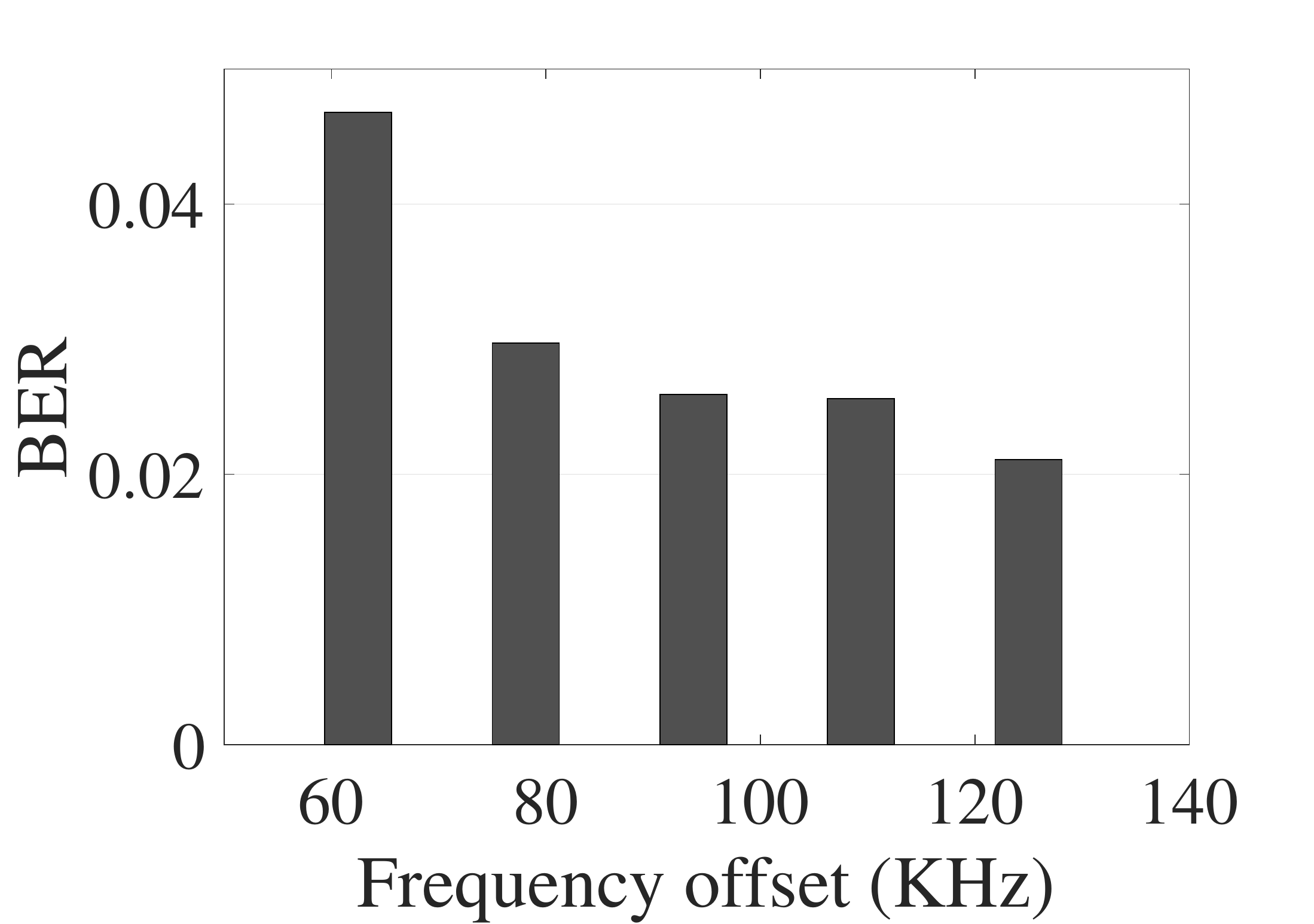}
		\label{ber_ble_freq}}
		\subfigure[BLE to Wi-Fi.]
		{\includegraphics[width=1.65in]{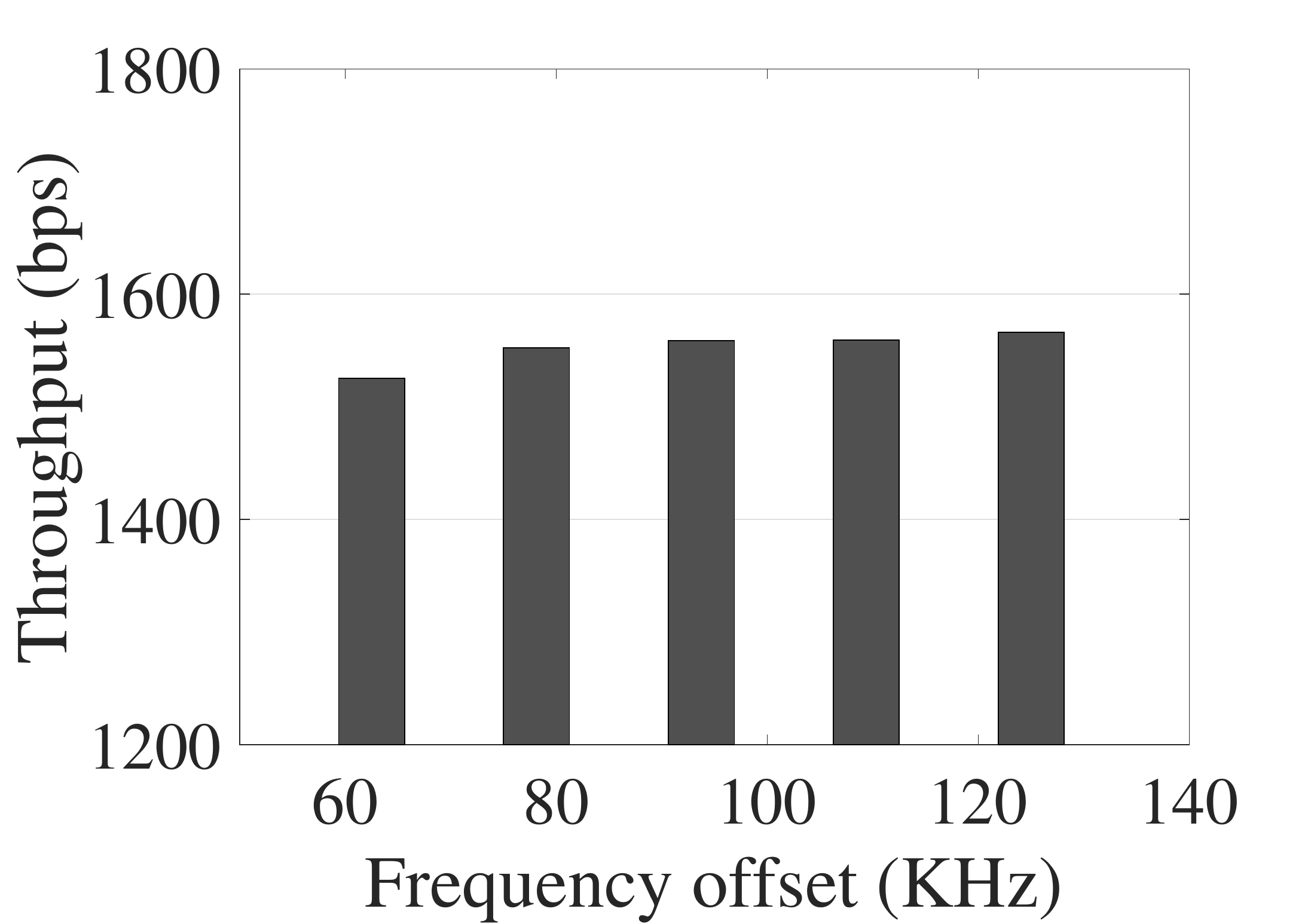}
		\label{th_ble_freq}}
		\caption{Impact of frequency shift on BLE to Wi-Fi links.}
		\label{fig:b2w_freq}
	\end{minipage}
\end{figure*}

\textbf{Impact of frequency shift.} Previous experiments use fixed artificial Doppler shift to embed bits. Now we study the impact of frequency shift on the performance of DopplerFi. Fig.~\ref{ber_wifi_freq} shows the impact of frequency on W2B links. The frequency offset in this experiment is shifted asymmetrically as the Wi-Fi preamble pattern received by BLE is asymmetric. We select 4 pairs of frequency shifts as depicted. As expected, BER decreases with the increment of frequency shift. We observe that with frequency shift larger than $\pm$100~KHz, W2B links yield BER as low as 0.12 and throughput higher than 250~bps. Recall that frequency shift no more than 150~KHz induces merely less than 0.8\% throughput loss to legacy Wi-Fi links. Thus, DopplerFi achieves a good balance by setting frequency shift in Wi-Fi senders within a range of 100-130~KHz.

The impacts of frequency shift on B2W links are plotted in Fig.~\ref{ber_ble_freq}, in which the BER of B2W links keep at a low level when the frequency shift is no less than 80~KHz. This is because the subcarrier spacing in Wi-Fi is 312.5~KHz, and the frequency difference must reach at least half of a subcarrier width to result in different patterns in CSI.

\textbf{Impact of overlapping channel.} Recall that BLE nodes adopt FHSS to transmit in different channels following a pseudo-random sequence. DopplerFi employs the DSK encoder to modulate DopplerFi bits in the overlapped channel. This set of experiments investigates the decoding ability of the CSI extractor when the Wi-Fi channel overlaps with different BLE channels. Regularly, there are 7 BLE channels overlapping with one Wi-Fi channel. Due to symmetry, there are totally 4 different overlapping patterns.

\begin{figure*}
	\centering
	\begin{minipage}[b]{3.5in}
		\subfigure[BER.]
		{\includegraphics[width=1.65in]{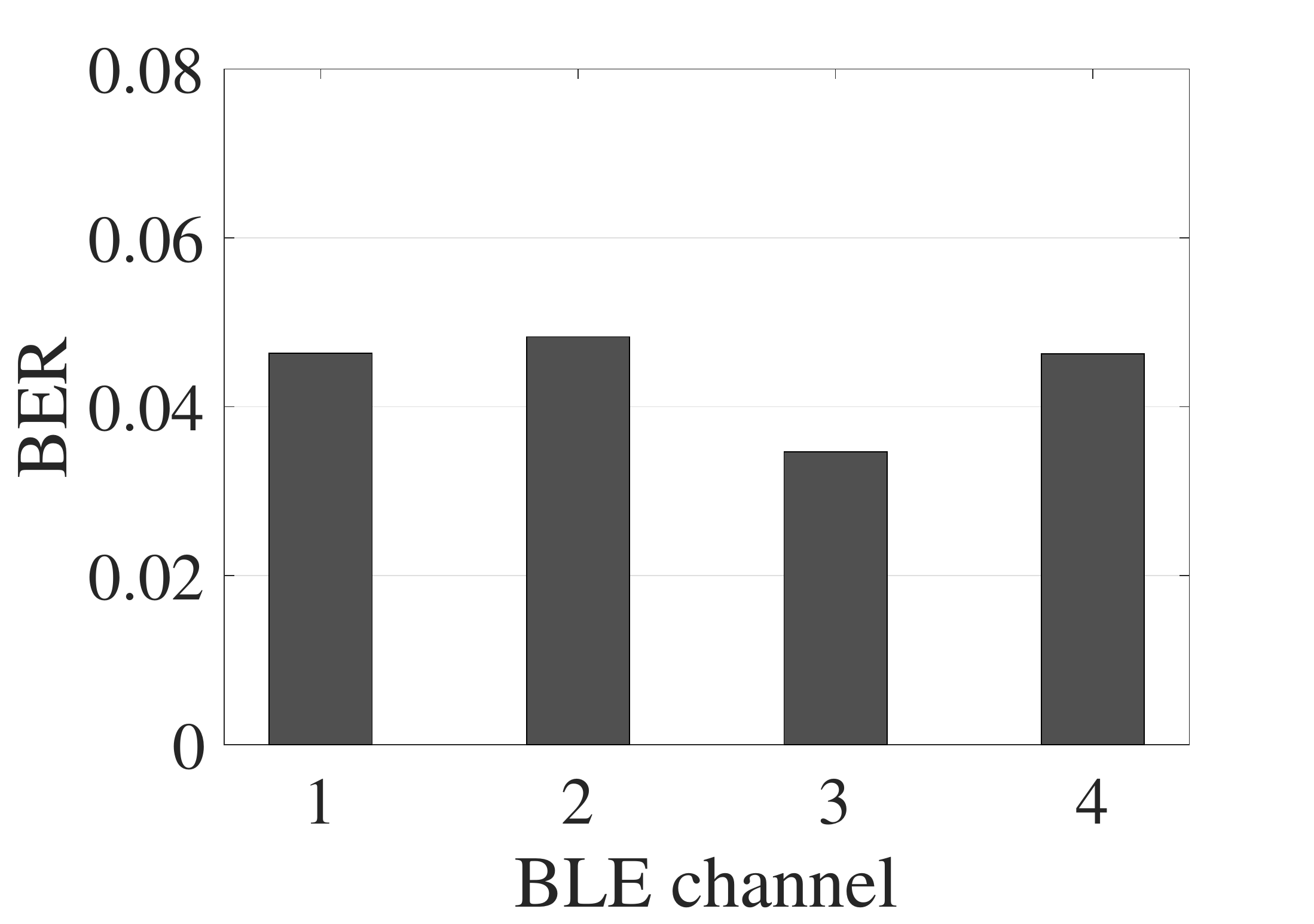}
		\label{ber_channel}}
		\subfigure[Throughput.]
		{\includegraphics[width=1.65in]{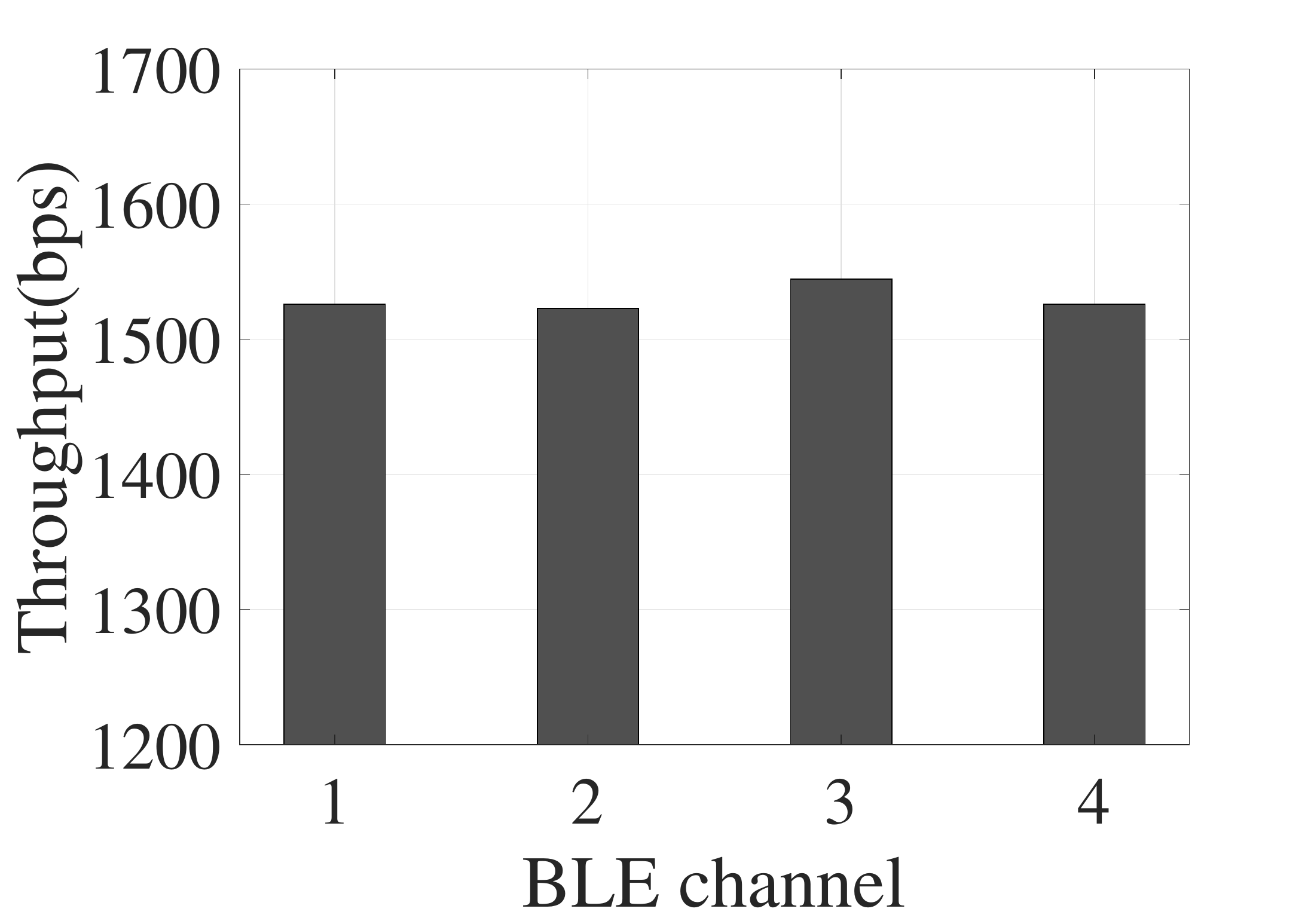}
		\label{th_channel}}
		\caption{Impact of BLE channel.}
		\label{fig:channel}
		\vspace{-0.2cm}
	\end{minipage}
	\begin{minipage}[b]{3.5in}
		\subfigure[BER.]
		{\includegraphics[width=1.65in]{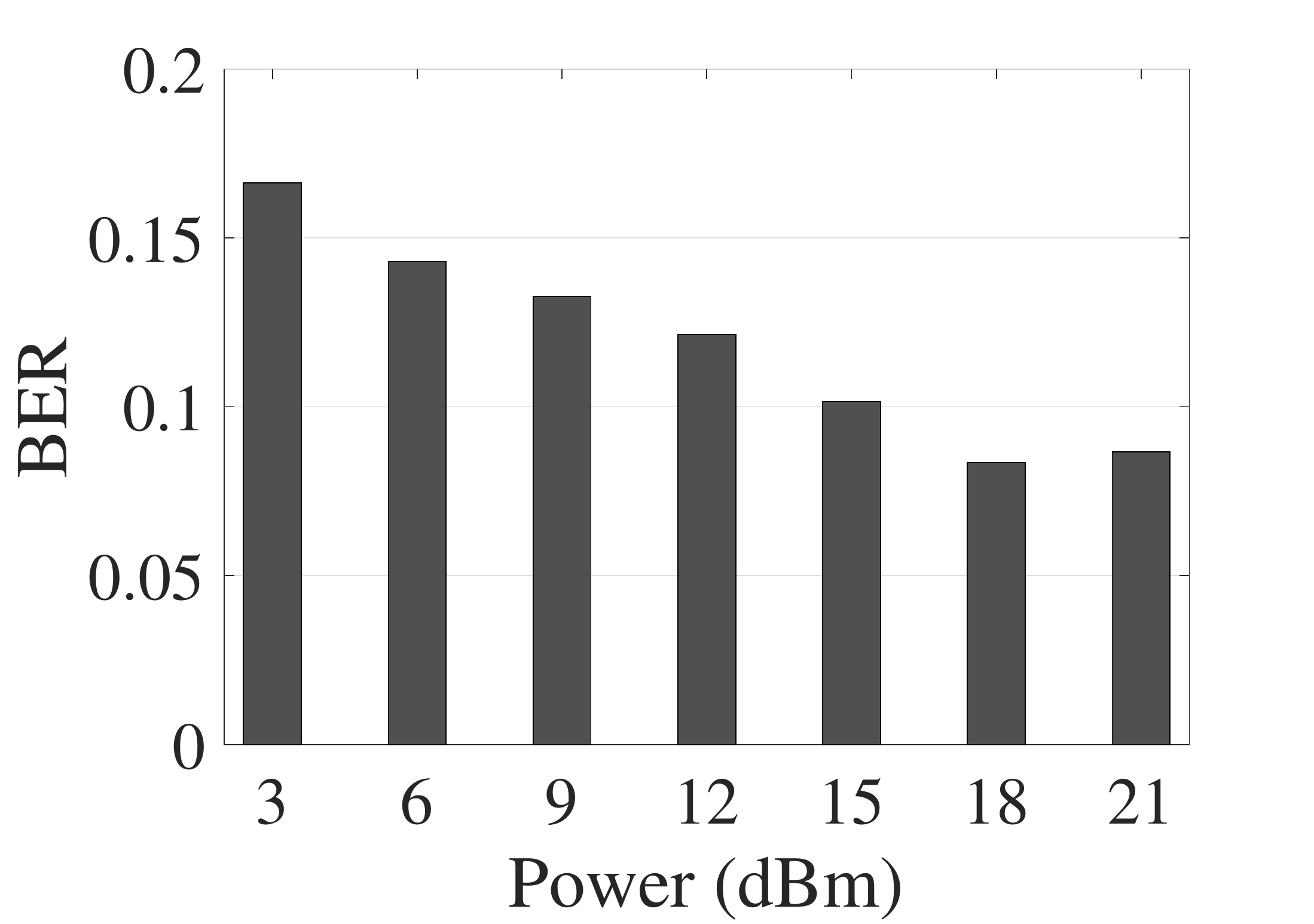}
		\label{ber_power}}
		\subfigure[Throughput.]
		{\includegraphics[width=1.65in]{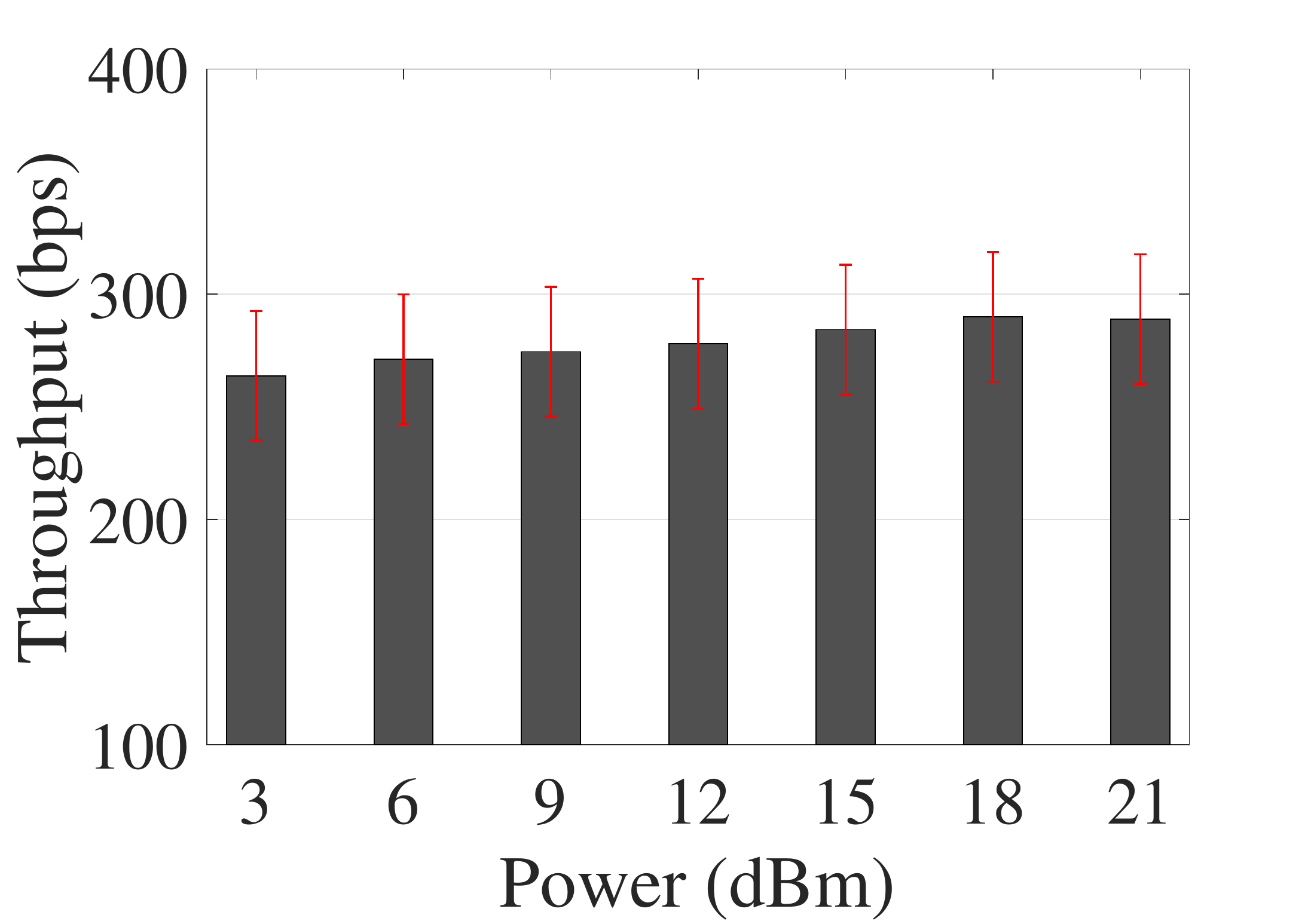}
		\label{th_power}}
		\caption{Impact of Wi-Fi TX power.}
		\label{fig:power}
		\vspace{-0.2cm}
	\end{minipage}
\end{figure*}

The Fig.~\ref{fig:channel} shows the BER and throughput performance under 4 different overlapping patterns. The Wi-Fi receiver collects samples in channel 1 and the BLE channels 1-4 indicate the channels from 2406~MHz to 2412~MHz. Although different patterns result in performance variance among multiple BLE channels, DopplerFi achieves BER of less than 5\% and throughput larger than 1.5~Kbps in all overlapping cases, which demonstrates that our demodulation scheme is robust in different overlapping patterns. 

\textbf{Impact of Tx power.} Next, we evaluate the impact of Wi-Fi's Tx power on DopplerFi. We use the same setting as in the distance experiments. At each location, we vary the transmit power from 3~dBm to 21~dBm by setting the power parameter in the 802.11 reference design of WARP.

Fig.~\ref{ber_power} compares the BER under various Tx power. DopplerFi experiences 16.63\% BER with Tx power of 3~dBm, and the BER drops to 8.67\% with Tx power of 21~dBm. With higher Tx power, DopplerFi receivers yield higher signal to noise ratio (SNR) and thus the GFSK patterns are more obvious. The Fig.~\ref{th_power} draws the throughput across different Tx power levels. The results show that although DopplerFi tends to achieve higher throughput with higher Tx power levels in most cases, the differences at high power conditions are marginal, which indicates that DopplerFi is robust even when the sender adjusts its Tx power using a power control mechanism.

\subsection{Link Coding Scheme}
\begin{figure}[t]
	\centering
	\subfigure[BER of B2W link.]
	{\includegraphics[width=1.6in]{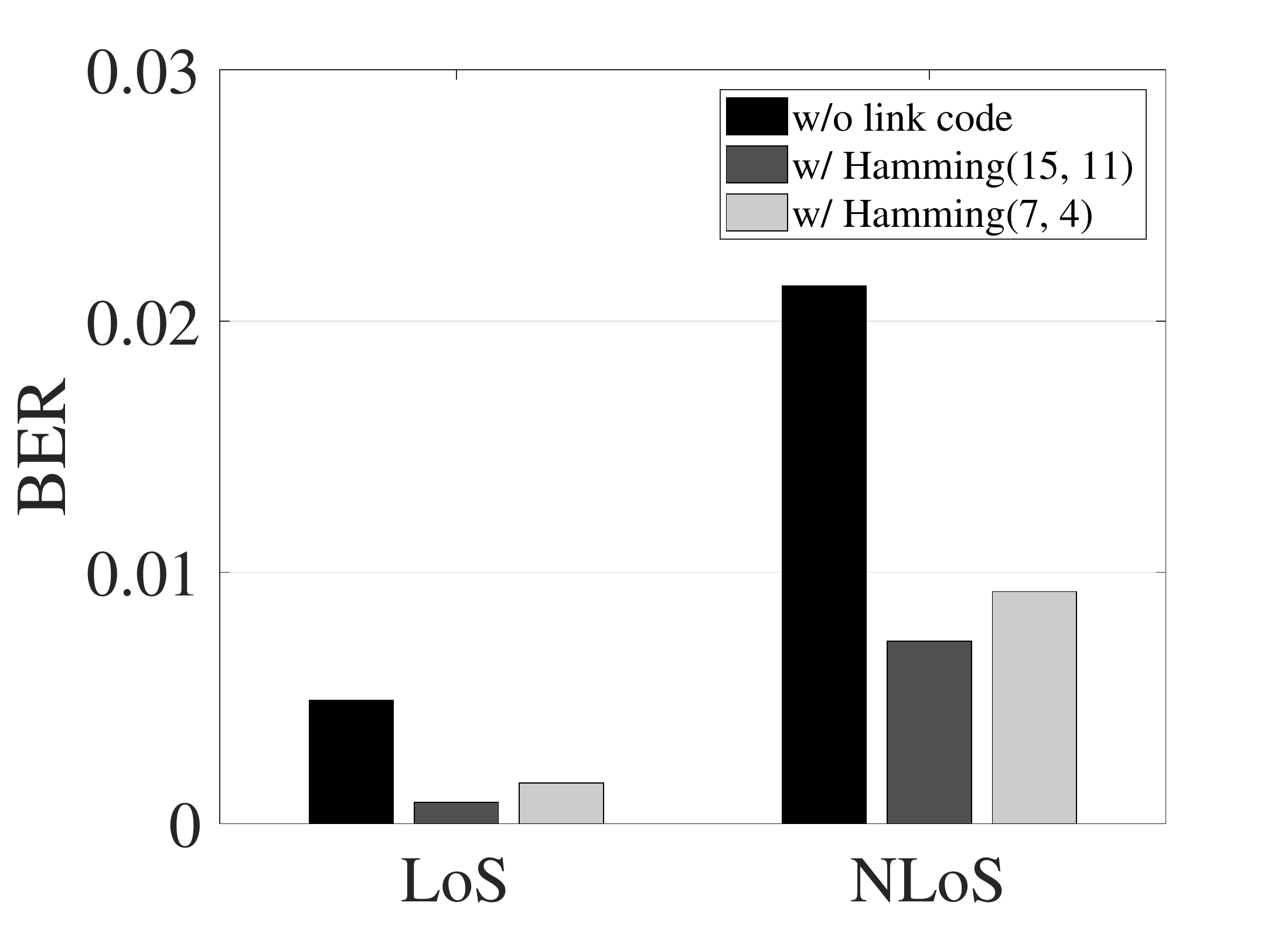}\label{c_ber1}}
	\subfigure[BER of W2B link.]
	{\hspace{0.1cm}
	\includegraphics[width=1.6in]{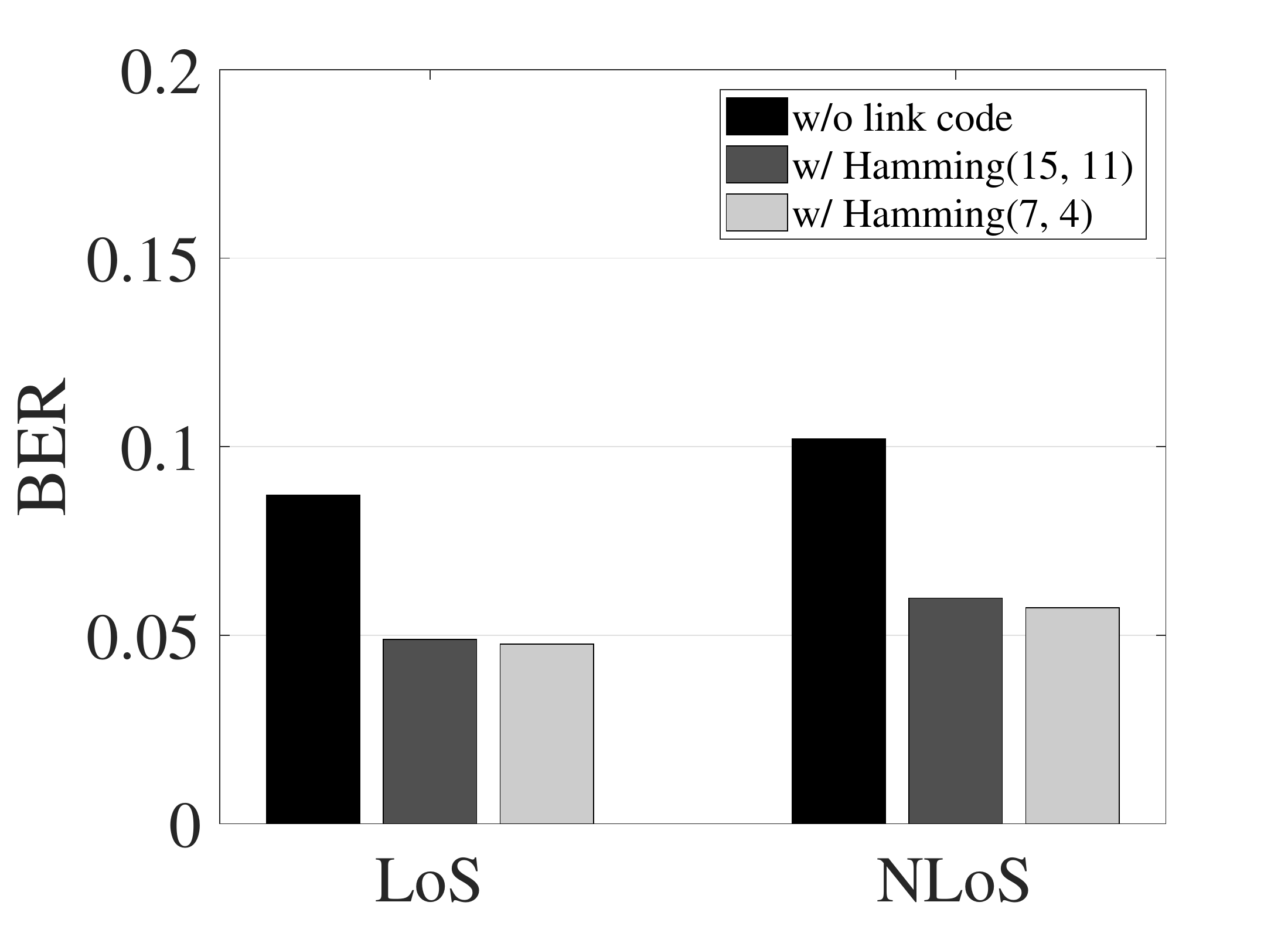}\label{c_ber2}}
	\caption{BER under Hamming coding}
	\label{fig:coding}
	\vspace{-0.2cm}
\end{figure}

Finally, we evaluate the DopplerFi system with two link coding schemes. Fig.~\ref{c_ber1} and Fig.~\ref{c_ber2} illustrate the performance of DopplerFi with link coding. Two types of link coding mechanisms are evaluated in this experiment for both B2W and W2B communication link.

We used Hamming code (15,11) and Hamming code (7,4) in our experiments. As the figures show, the error of some transmitted bits can be recovered, which reduces the symbol error of DopplerFi. When the link coding scheme is used, the BER of W2B link decreases to 4.9\% in LoS scenario and 5.8\% in NLoS scenario at close distance, respectively. For B2W link, the BER of Hamming code is lower than 0.1\% when the Hamming code is used in LoS scenario and lower than 1\% in NLoS scenario. 

Comparing to the BER without link coding in LoS scenario, 43.74\% of the corrupted bits have been recovered with Hamming code (7, 4) in W2B link. Similarly, 83.7\% of the error bits have been recovered in B2W link. The figures also show that the difference between the two coding schemes is very small if the uncoded BER is low, and the difference would slightly increase as the BER become larger. Due to the higher BER in the W2B link, the Hamming coding scheme produces a larger improvement to recover the error.

\section{Related Work}\label{sec:literature}

The design of DopplerFi is inspired by CFO related side channels~\cite{carpool,wang2016privacy}, which exploit CFO-induced phase rotation in OFDM systems to facilitate different functionalities. Carpool~\cite{carpool} injects extra phase shift in each OFDM symbol to create a free side channel in existing OFDM PHY structure to carry symbol-specific parity check bits. In contrast, PriLA~\cite{wang2016privacy} injects excessive CFO that varies across OFDM symbols to prevent Wi-Fi eavesdroppers from correctly decoding packets. The CFO pattern is generated from a CSI key shared by the sender and the receiver to make the CFO recoverable at the receiver. Different from these proposals, DopplerFi studies a CFO-based side channel for cross-technology communications. Technically, DopplerFi departures from existing CFO-based side channels in that existing works utilize the inherent CFO estimation block in OFDM PHY for demodulation, while DopplerFi makes it possible to demodulate CFO changes using incompatible PHY.

Interference avoidance and cancellation through cross-technology interaction has been investigated in many previous works~\cite{wizbee, cbt, rfsmog, zimo, lte_wifi,zheng2017interference}. WizBee~\cite{wizbee} leverages decoded data for channel estimation to extract ZigBee signals in the presence of strong interference from 802.11 nodes. 
CBT~\cite{cbt} enables the reliable coexistence between Wi-Fi and ZigBee network by sending a busy tone along with the data transmission. TIMO~\cite{rfsmog} enables 802.11n equipped with MIMO design to communicate in the presence of high-power cross-technology interference. Differently, ZIMO~\cite{zimo} leverages the channel coefficient of the ZigBee and Wi-Fi technologies, which handles both signals at the same time. The proportional fair rate allocation~\cite{lte_wifi} is derived to allow the coexistence of random access and scheduled transmitters between LTE and Wi-Fi. ZiSense~\cite{zheng2017interference} extracts time-domain features in RSSI in the presence of coexisting interference. Different from these approaches, DopplerFi aims to build a direct communication link cross different protocols.

DopplerFi belongs to the category of cross-technology communication. To save the energy of Wi-Fi radios, ZiFi~\cite{zifi} leverages ZigBee radios to identify the existence of Wi-Fi networks through interference signatures caused by Wi-Fi beacons. WizSync~\cite{wizsync} utilizes the periodic Wi-Fi beacons to calibrate the clocks of ZigBee nodes. WizNet~\cite{WizNet} detects Wi-Fi signals from RSSI measured by ZigBee automatically. Direction communications between incompatible PHYs are studied in~\cite{esense,howies,gsense,freebee,b2w2,cmorse,DCTC,EMF,wizig}. Recent advances~\cite{freebee,b2w2,cmorse,DCTC,EMF,wizig} explore more transparent approaches to embed side-channel bits without modifying the data of legacy packets. Differently, our goal is to explore the frequency dimension that has minimal impact on MAC behaviors.

\section{Discussion}
\label{sec:discuss}

\subsection{Gateway}
BLE nodes are conventionally controlled by a central gateway. Similarly, Wi-Fi clients connect to the Internet through a Wi-Fi AP. DopplerFi does not make any assumption or change on BLE or Wi-Fi topologies, but creates an extra link across BLE and Wi-Fi nodes. A traditional solution to allow BLE nodes to connect to the Internet is to deploy a dedicated BLE-Wi-Fi gateway that converts BLE signals to Wi-Fi-compatible format. The goal of DopplerFi is to eliminate such a type of gateway and enable ubiquitous communications between BLE and Wi-Fi without extra deployments.

\subsection{Optimal Parameters}

In our experiments, the thresholds for GFSK detection and demapping are empirically set. In particular, these values are empirically set based on the sample distributions observed in our experiments to achieve optimal detection and decoding performance. To derive the optimal threshold, we need to identify the impact of the channel and noise in the current environment. Since we do not make any assumptions about the channel model, interference, and noise in our design, it is prohibitive hard to determine the relation between the parameters and the environments, which require complete information about the current channel state and the environment. Since the focus of this paper is to propose a lightweight and practical design with minimal assumptions about the environments and the networks, we tend to discuss the parameter optimization problem in future work.

\section{Conclusion}
\label{sec:conclude}

This paper introduces DopplerFi, a cross-technology communication framework that aims to introduce minimal disturbance to legacy networks. By exploiting the redundancy in carrier frequency shifts, DopplerFi establishes a free side channel without modifying transmission power or time of legacy packets. While our current implementation of DopplerFi demonstrates the bidirectional communications between Wi-Fi and BLE, we believe the framework can be extended to support other wireless technologies.

\bibliographystyle{IEEEtran}
\bibliography{IEEEabrv,./WiFiBLE}

\begin{IEEEbiography}[{\includegraphics[width=1in,height=1.25in,clip,keepaspectratio]{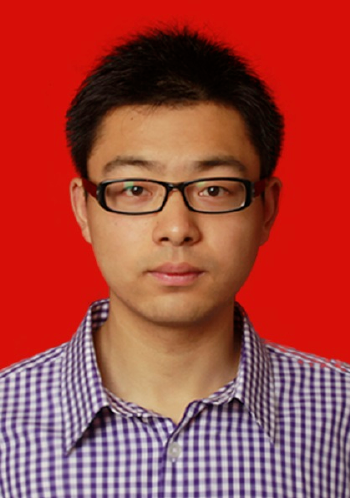}}]{Wei Wang (S'10-M'16)} received the Ph.D. degree from the Department of Computer Science and Engineering, The Hong Kong University of Science and Technology. He is currently a Professor with the School of Electronic Information and Communications, Huazhong University of Science and Technology. His research interests include PHY/MAC design and mobile computing in wireless systems. He served on TPC of INFOCOM and GBLOBECOM. He served as Editors for IJCS, China Communications, and Guest Editors for Wireless Communications and Mobile Computing and the IEEE COMSOC MMTC COMMUNICATIONS.

\end{IEEEbiography}

\begin{IEEEbiography}[{\includegraphics[width=1in,height=1.25in,clip,keepaspectratio]{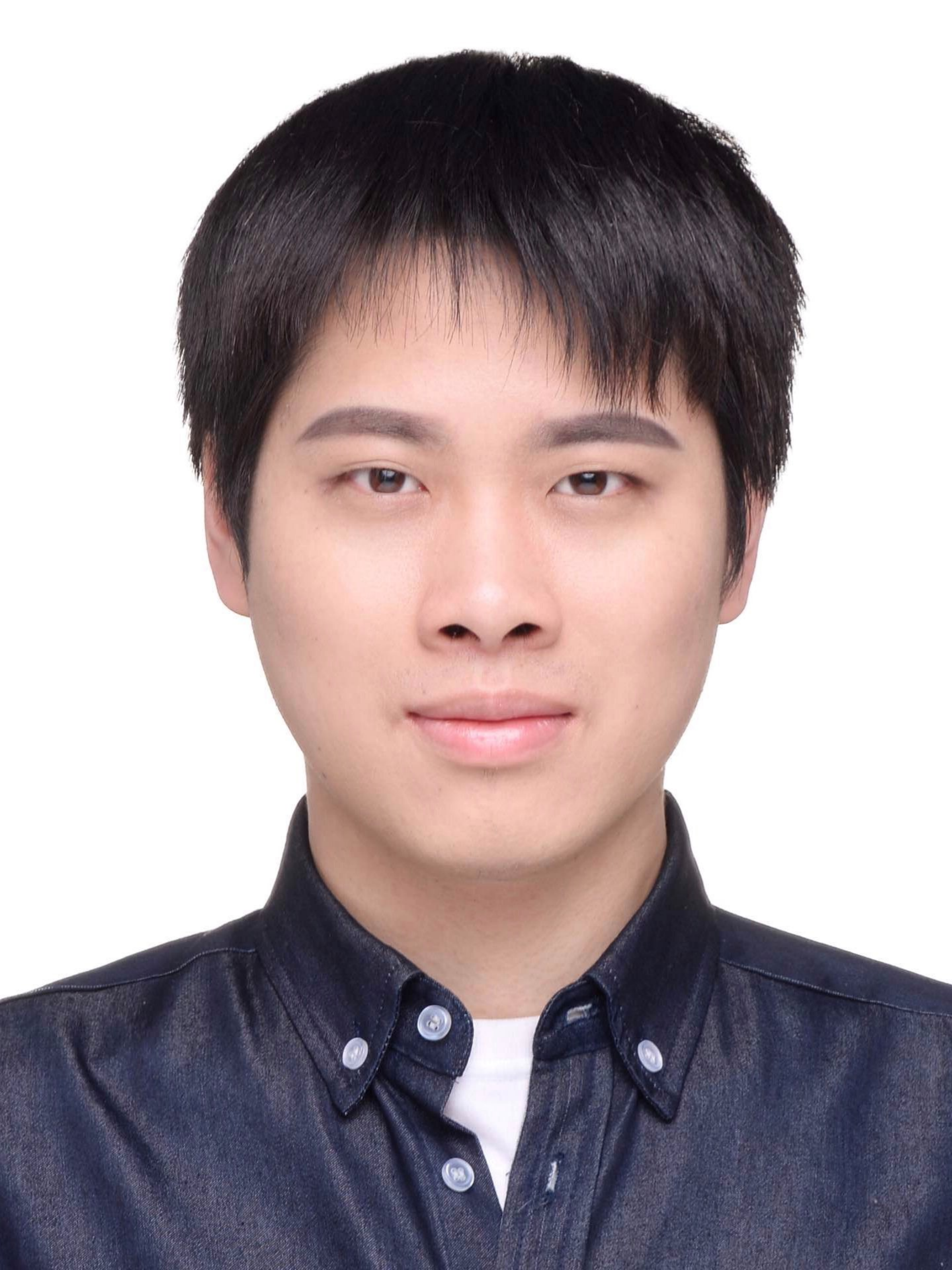}}]{Shiyue He} is currently pursuing his Ph.D. degree at School of Electronics and Information Engineering, Huazhong University of Science and Technology, Hubei, China. Before that, he has received his Bachelors degree in information engineering from Wuhan University of technology, Hubei, China, in June 2016. His research interests include PHY/MAC communications and sensing in wireless networks.
\end{IEEEbiography}

\begin{IEEEbiography}
[{\includegraphics[width=1in,height=1.25in,clip,keepaspectratio]{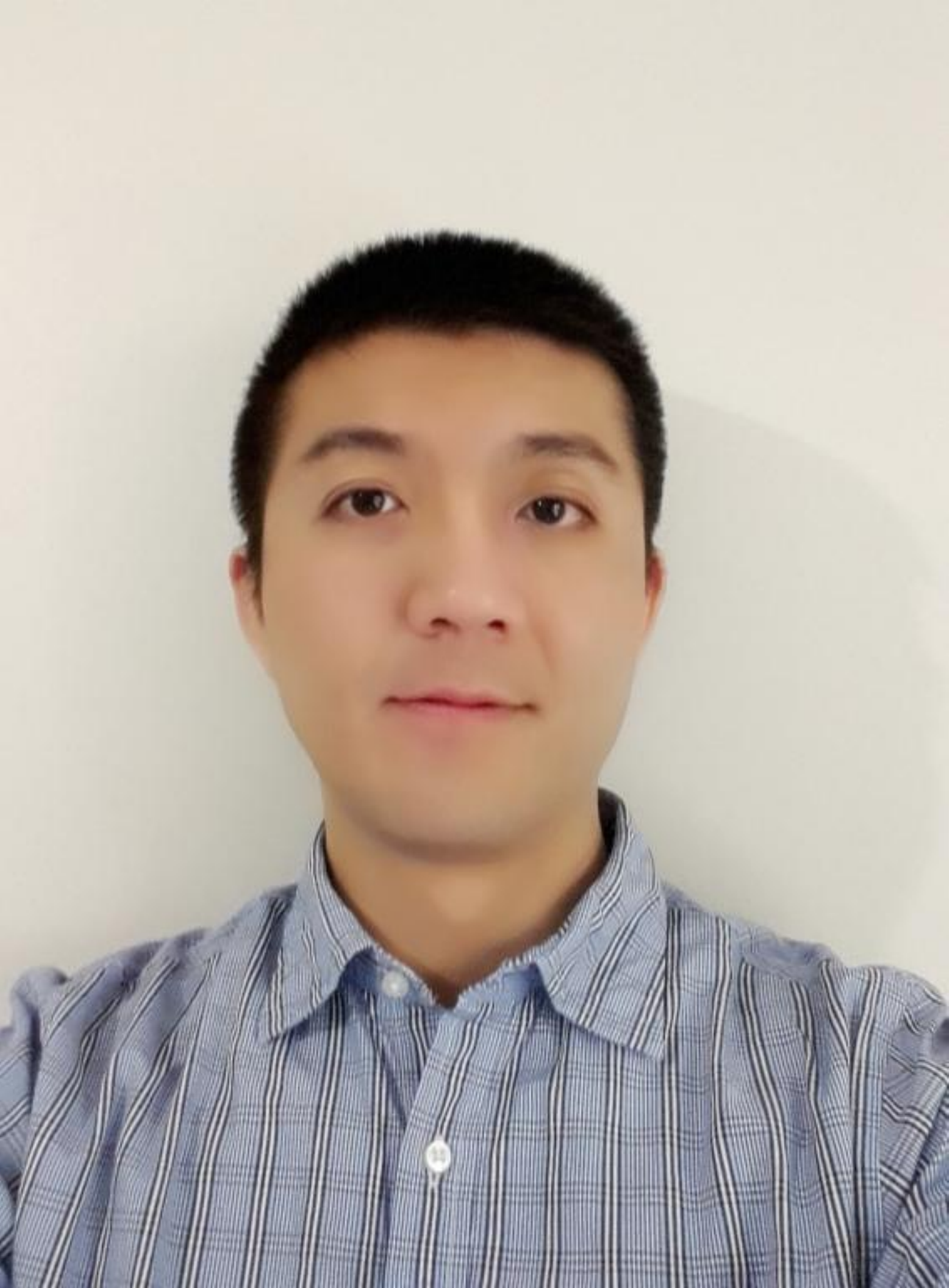}}]{Liang Sun (S'08-M'12)} received Ph.D. degree in Electrical Engineering from the Hong Kong University of Science and Technology in 2010. From 2010 to 2014, he was a researcher with the Alcatel-Lucent Shanghai Bell and the NEC Corporation. From 2014 to 2015, he had been a postdoctoral research fellow with the Department of Electrical and Computer Engineering, the University of British Columbia, Vancouver, BC, Canada. Since 2016, he has joined the Beihang University as an associate professor. His research interests include information theory and signal processing. Dr. Sun received a 2010 Young Author Best Paper Award by the IEEE Signal Processing Society.
\end{IEEEbiography}

\begin{IEEEbiography}[{\includegraphics[width=1in,height=1.25in,clip,keepaspectratio]{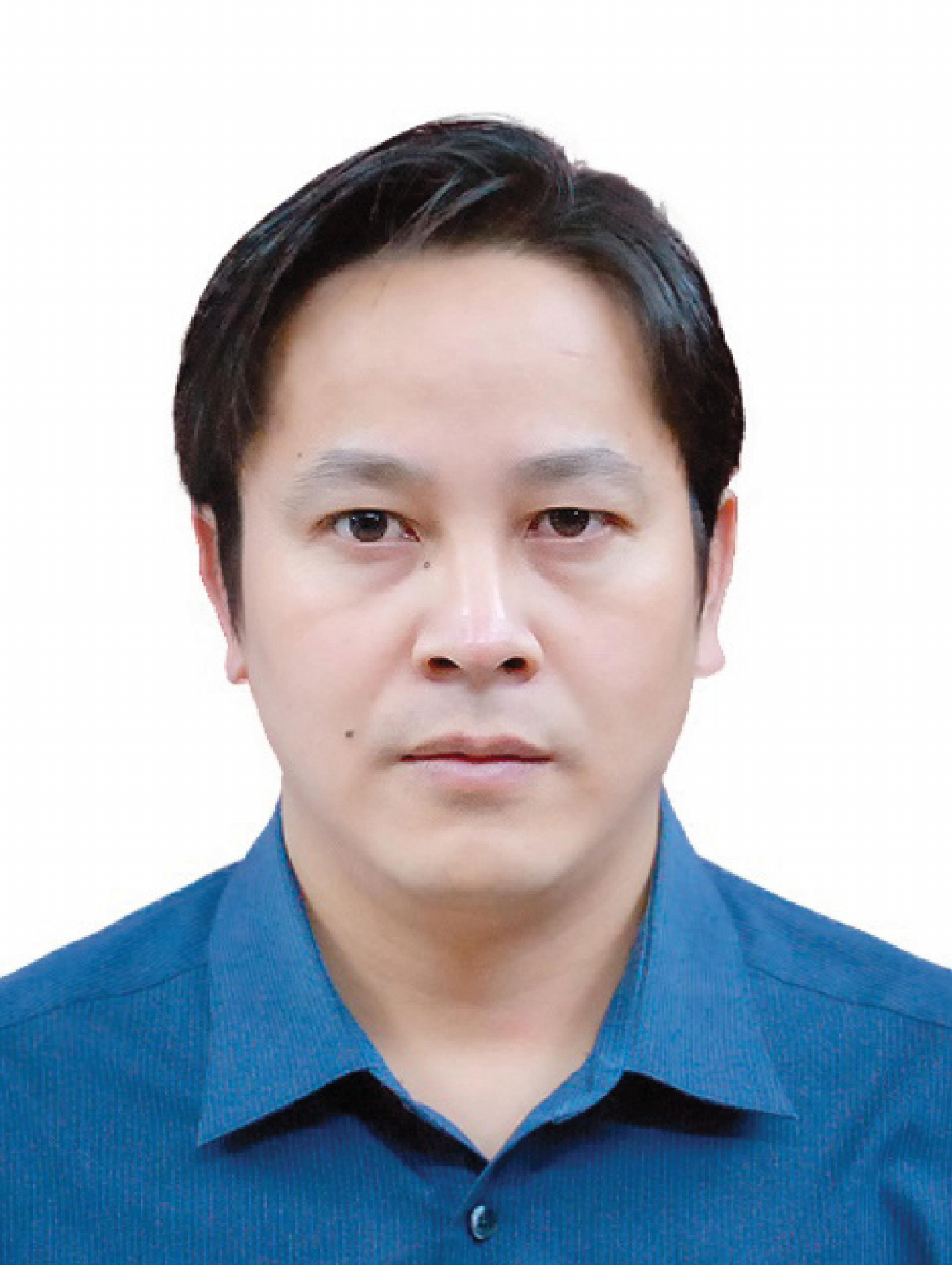}}]{Tao Jiang (M'06-SM'10-F'19)} is currently a Distinguished Professor in Huazhong University of Science and Technology (HUST), Wuhan, P. R. China. He received Ph.D. degree in information and communication engineering from HUST in 2004. He worked in Brunel University and University of Michigan-Dearborn, respectively. He served as associate editors in IEEE Transactions on Signal Processing, IEEE Communications Surveys and Tutorials, and is the associate editor-in-chief of China Communications. He is a recipient of the NSFC for Distinguished Young Scholars Award in 2013， and he is a Fellow of IEEE.
\end{IEEEbiography}

\begin{IEEEbiography}[{\includegraphics[width=1in,height=1.25in,clip,keepaspectratio]{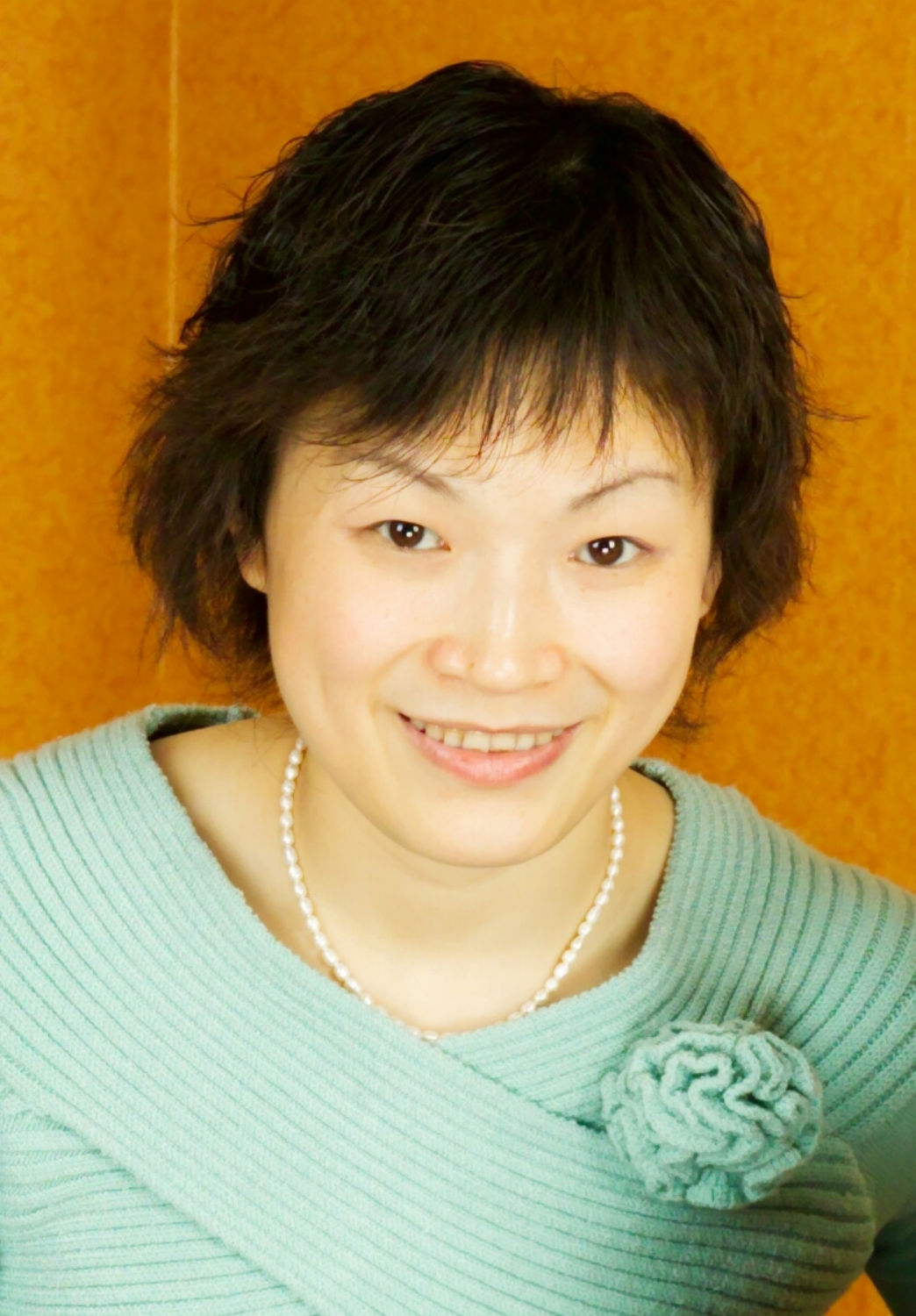}}]{Qian Zhang (M'00-SM'04-F'12)} joined Hong Kong University of Science and Technology in Sept. 2005 where she is a full Professor in the Department of Computer Science and Engineering. Before that, she was in Microsoft Research Asia, Beijing, from July 1999, where she was the research manager of the Wireless and Networking Group. She is a Fellow of IEEE for ``contribution to the mobility and spectrum management of wireless networks and mobile communications". Dr. Zhang received the B.S., M.S., and Ph.D. degrees from Wuhan University, China, in 1994, 1996, and 1999, respectively, all in computer science.

\end{IEEEbiography}

\end{document}